# Apparent Spatial Memory Register and Oscillating Shape Transformations in Gold Nanostructures

Sudhir Kumar Sharma,[1] Renu Pasricha,[2] James Weston,[2] Thomas Blanton[3] and Ramesh Jagannathan[1*]

[1] Engineering Division, New York University Abu Dhabi, Abu Dhabi, UAE.

[2] Core Technology Platform, New York University Abu Dhabi, Abu Dhabi, UAE.

[3] International Centre for Diffraction Data, 12 Campus Blvd., Newtown Square, PA19073

*Correspondence to: Ramesh Jagannathan (Email: rj31@nyu.edu )

**Abstract:**

Nano-structures, namely, nano-rings and nano-particles, comprising two-dimensional, crystalline gold atomic discs were produced on a sapphire substrate. Nano-rings revealed several remarkable properties, namely, an apparent spatial/orientation memory register, an apparent programmed "homing instinct" and "*action*" based on that instinct, usually observed only in **living, biological** systems. Momentary application of 100nN force on **a** nanostructure ('mother') led every 'mother' within the entire scanner area to increase its volume and create several "daughter" replicas. Over time, *without* the influence of any applied external force or stimulus, each 'daughter', of its own volition, transported itself back to its *'mother'* and re-created its original form. The results strongly implied temporal and spatial synchronization between the 'daughters' and their respective

'mothers'. The targeted movement of the 'daughters' towards their 'mothers' implied the existence of a special, yet unidentified new force. Momentary application of a weak force on *a* gold nanoparticle resulted in the spontaneous transformation of each nanoparticle into a complex and 'robot-like' (e.g. Dalek) 3-dimensional form. Without the application of any additional external force or stimulus, we observed a remarkable display of oscillatory transformation between the 'robot-like' forms and their 'mother' nano-particles lasting over 26 hours. The "robot-like" structures seem to replicate identical copies of themselves.

**Introduction:**

Two dimensional materials (2DM) are one atom/molecule thick crystalline planar structures and Smart Materials (SM), are defined primarily as those whose one or more properties are tunable in a controllable and reversible fashion by an applied external stimulus. These are two of the most active fields of research, because of the remarkable properties of such materials and the promise they hold together in terms of applications across varied disciplines. [1–9] Our research was specifically focused on developing a technique to prepare 2D atomic structures of gold, with a focus to complement and extend the advances in the field of metal nanostructures. [8–11] Our starting interest in gold 2DM was due to its potential high impact applications in fields such as photonics, electronics, optical-sensing/imaging, and drug-delivery. [12,13] While researchers in the field have developed a range of gold nanostructures using various biological, photochemical and wet chemical methods. [6,14–18], so far little progress has been made in the synthesis and characterization of gold 2DM. The results of our experiments, however, led to unexpected and remarkable discoveries.

**Results:**

The hypothesis behind our experimental procedure was based on the well-known phenomenon that the surface melting point of a thin film is significantly lower than its bulk melting point. In our experiments, a thin film of gold (e.g. X-ray Reflectivity (XRR) thickness ~ 66 nm (Figure S1) was deposited on a single crystal sapphire substrate. The above sample, when exposed to a temperature of 475 $^0$C for a short duration of time (t = 5 – 60 seconds), that is significantly lower than its bulk melting point (1064 $^0$C), could result in the following phenomenon: the gold film in contact with sapphire would start to melt and recede from the interface; it would instantly become supersaturated and be forced to precipitate back on the substrate which would, once again, trigger its melting; this would create an untenable, cyclical process of melting and re-precipitation; the only realistic solution out of this quagmire would be if the re-precipitation of gold on the sapphire resulted in a 2D structure of gold (which, by definition, does not have any "bulk") with a preference to grow perpendicular to the substrate.

Our experiments produced both nano-ring structures and nano-particles. Nano-rings were observed for the 66 nm (XRR) thick samples that were treated to 475$^0$C, for 2 - 60s, respectively. AFM images of nano-ring structures observed for the 2s, 5s, 30s and 60s heat-treated samples are shown in Figure S2. The AFM instrument we used is capable of operating in a "true" non-contact mode to facilitate non-invasive image capture. We were unable to effectively image the samples using the SEM, due to significant beam damage (Figure S3). Since sapphire substrate was used in our experiments, TEM was ruled out as a primary method to study the phenomenon described below. However, TEM was effectively used for chemical and structural characterization. The results from nano-rings experiments reported in this manuscript are for samples, which were treated to $475^0$C for 60s, unless stated otherwise. The results from nano-particles experiments (Figure S4). reported in this manuscript are from 200 nm thick (QCM) gold coatings on sapphire treated to 550 $^0$C for 30 seconds.

**Nano-ring Structures:**

*A. Characterization of Nano-Structures*

As per our hypothesis, we observed circular, and "open top" dome like growth structures of gold film, rising from their edges that were pinned to the substrate, as shown in Figure 1A and Figure S5. Over time, these structures appear to coalesce with each other to grow stacks of larger circular arcs, eventually leading to well-formed asymmetric nano-rings (Figure 1B-F, Figure S6). High-resolution AFM image of a selected area of the nano-ring (Figure 1G - white arrow) is shown in Figure 1H. It is evident that the nano-ring is a composite assembly comprising several smaller discs, as their basic building blocks. In Figure 1I, we measured the height profile across these smaller discs to be approximately 1.7 angstrom, implying one atomic layer of gold. [8,19–21]

X-ray diffraction (XRD) confirmed the presence of gold with (200) and (220) diffraction peaks observed (Figure S7) in the XRD pattern. Deposition of gold films typically results in a strong (111) diffraction peak (~38.5 degrees two-theta using Cu Kα radiation). Absence of the gold (111) diffraction peak in our samples would imply a significant change in the morphology of the gold due to the $475^0$C, 60s thermal processing.

High-resolution AFM phase image of a typical nano-ring (Figure 1G) clearly showed that each is composed of several smaller discs (Figure 2A). The base of the nano-rings on the substrate was also found to be covered with similar, smaller disc structures. STEM imaging confirmed the AFM results (Figure 2B) and that the nano-ring is indeed gold (Figure 2C, F and G). High resolution TEM (HREM) (Figure 2D and E) confirmed a gold lattice spacing of 2.8 Å. It has already been reported that for films with less than three atomic layers, one could see the (110) forbidden reflection for gold with a lattice spacing of 2.8 Å. [22–24] TEM diffraction patterns, confirmed the presence of 2-dimensional gold discs of atomic thickness in our samples.

### B. Spatial Memory Register and Homing Instinct

At room temperature, we subjected one of the nano-rings to a momentary weak force (100 nN) using the Force/Displacement spectroscopy technique in the Atomic Force Microscope [25–27]. The force was applied at selected locations, shown by the red ink **(+)** marks in the image in Figure 3A. We show the AFM scan collected for the same sample field before (Figure 3A) and after the application of the force (Figure 3B-D), at various time intervals. The application of force resulted in a significant increase in volume for each (nano-ring) 'mother' in the entire scanner area and formation of several 'daughter' replicas per 'mother'. They were found to be scattered around and away from their respective 'mothers' (Figure 3B). Over a period of time, without the application of any additional external force or stimulus, each 'daughter', of its own volition, returned to its respective 'mother' and re-created the original ('mother') form and size (Figure 3C and 3D). The final AFM image of the 'mothers' (Figure 3D) was identical to that observed prior to the application of the force (Figure 3A). It is impressive that each 'daughter' ONLY returned to its respective 'mother' retaining each 'mother's' original orientation fidelity during the re-assembly process. It is important to emphasize that even if two 'daughters' from two different 'mothers' overlapped and were in significant physical contact with each other, they always returned to their respective 'mothers' (Figure S8). In our experiments, we did not encounter a situation where some of the 'daughters' had completed the navigation process and re-assembly, while others have not.

The effect of the force on the nano-ring shown in Figure 3A was simultaneously and uniformly experienced by all the nano-rings within the 100 μm x 100 μm scanner bed. In Figure S9, we show that nano-rings located 45μm away from the point of application of the force exhibiting the same series of behavior as shown in Figure 3. In each experiment, we had imaged several nano-ring structures across the entire scanner area and found all of them to exhibit this phenomenon. In another example, we show two sets of nano-rings, one in Figure S10A (x=3μm, y=24.6μm) and a second one in Figure S10D (x=3.3μm, y= -20.55μm), which was 45μm away from the first set. We applied a momentary force of 200nN to the sapphire substrate in locations marked by the white ink (+) sign in Figure S10A. This resulted in the creation of 'daughters' from each "mother" in both locations (Figure S10B and S10E). After a few hours, all the 'daughters' returned to their respective 'mothers' in unison (Figure S10C and Figure S10F), consistent with our earlier findings.

### C. Coherence Phenomenon

We were able to activate following phenomenon by exposing the nano-rings to acoustic waves and the results unequivocally confirmed strong coherence between the nano-rings. We subjected the nano-rings (Figure 4A) to an acoustic frequency of 10 KHz for 30 seconds. It resulted in the creation of one 'daughter' from each 'mother'. The relative direction, location/orientation, shape and size between the "mother/daughter" pair was identical for all the imaged structures (Figure 4B). 49 minutes later, all the 'daughters' had returned to their respective 'mothers' in unison and a second 'daughter' was generated on the opposite side of each "mother" (Figure 4C). Once again, the relative direction, location/orientation, shape and size between the "mother/daughter" pairs were identical for all the imaged structures. After 126 minutes, all the second generation 'daughters' returned to their respective 'mothers' in unison. (Figure 4D). It is evident that the first and second generation 'daughters' were ***not*** in physical contact with their 'mothers'. In order for the observed phenomenon to occur, it required the following: temporal and spatial synchronization between the 'daughters' as they navigated back to their respective 'mothers', precise directional, spatial and shape register in each 'daughter' regarding its genesis (i.e. "mother"), a dynamic communication/navigational control feedback loop between all the 'mother'/'daughter' pairs and a governing force field. This was a remarkable observation with significant implications.

### D. Synchronized Shape Transformation Phenomenon

We report on the shape transformation phenomenon exhibited by nano-ring structures when subjected to a vibration frequency of 125 Hz for 10 seconds. Figure S11A shows the image of the nano-rings before exposure. Figure S11B shows the image of the same field immediately after exposure. Scanning from "bottom to top", the bottom half of the scan showed that each nano-ring had transformed into a larger truncated "cone-like" structure due to the vibrational waves. Half-way through the scan, we observed a highly synchronized, sudden transformation of all the truncated "cone-like" structures into nano-rings. Subsequent scan revealed only nano-ring structures (Figure S11C). Figure S11D shows the transformation of nano-rings back to the truncated "cone-like" structure, after 27 hours. An AFM image of the clean sapphire substrate, for reference purposes, is shown in Figure S12.

**Nano-Particle Structures:**

*E. Volume/Shape Transformations*

Our experiments with nano-particles were similar to those carried out with the nano-rings. In several instances, the nano-particles increased their volume significantly (Figure 5) as a first response to an applied force and returned to their original state after a given time interval. Figure 5A shows nano-particles prior to application of a 100 nN force on the red (+) mark location. Subsequent scan shown in Figure 5B shows a significant increase in volume of all the nanoparticles. Figure 5C shows the return back to the original nanoparticle size/form after a period of 24 hours. In some cases, there was no correlation between the shape/form of the nanoparticle before and immediately after the application of the force. For example, a circular hockey puck shape (Figure 5D) transformed into a highly irregular one (Figure 5E) accompanied by a very large increase in volume (~100X). Please see the scale markers in Figure 5D and 5E. In many cases, we found that the increase in volume was also accompanied by each structure growing identical and significant additional features. For example, in Figure S13 we show that after increasing its volume due to applied force (Figure S13A), each and every 'mother' created one 'daughter' on a specific corner location (Figure S13B) that moved over time towards the center (Figure S13C). Subsequent scans revealed a pair of distinctive twins ('daughters') centrally located on each 'mother' (Figure S13D). After 3 hours, each 'mother' lost one 'daughter' (Figure S13E) and soon the original structure/form re-appeared (Figure S13F). All of these collective transformations appear to be highly synchronized.

*F. Robot-like (e.g. Dalek) Structures and Oscillatory Shape Transformation Phenomenon*

In another experiment, momentary application of a 100 nN force on a gold nanoparticle resulted in a spontaneous transformation. Each nanoparticle transformed into a larger and a highly complex 3-dimensional "robot-like" (e.g. Dalek) form.  We applied a force on the location marked by the white ink (+) mark (Figure 6A). The resultant "robot-like" structures had an uncanny resemblance to each other. (Figure 6B). We observed a remarkable phenomenological display of, for lack of a better word, an "oscillatory dance" between these "robot-like" forms and their original "mother" nano-particles, in the absence an external force or stimulus. It is important note that the phenomenon was still ongoing even after 26 hours (Figure 6C and 6D). To validate this, we show the intermediate recorded evidence of this 'oscillatory dance' during the 26 hours period (Figure S14).  In Figure 6E (Figure S15), we show the higher magnification images of these "robot-like" structures which are  *replica copies* (Figure 6F and Figure S15).  We also observed the transformation phenomenon occurring several hours earlier at a location that is 33.65 μm away (Figure S16) from the marked location in Figure 6.

In general, the nano-particles responded to an applied force by creating various highly complex and novel 3-dimensional structures/forms that exhibited this phenomenon (Figures S17, S18 and S19). For example, application of the force led each 'mother' (Figure S17A) to create an adjacent 'daughter' (Figure S17B) and all them were precisely oriented at a specific angle.  Over time, each 'mother/daughter' pair transformed into an identical, significantly larger 3-dimensional structure (Figure S17C). Subsequently, each of these structures transformed into a new well-defined form (Figure S17D). They were highly ordered and oriented in a specific angle that is a mirror of the angle observed in Figure S17B. After several hours, these structures transformed into yet another new form, with no particular directional order or orientation (Figure S17E). These structures then returned back to their original nano-particle form (Figure S17F) with a mirror orientation of Figure S17B. Further confirmation of this phenomenon is evidenced in Figures S18 and S19. Figure S18 shows 'robot-like' structures that have an uncanny resemblance to a typical 'human-like' form. Images in Figure S19 have a striking resemblance to an 'animal-like' form. In Figure S20, we show examples of other complex 3-dimensional structures formed by nano-particles as a force response.

**Discussion**

We repeatedly observed that, after being subjected to a momentary weak force, nano-ring 'mother' structures undergo the process of creating 'daughter' structures that are replicas of their respective 'mothers'. Over time, the 'daughters', without fail, navigated back to their original 'mothers'. To illustrate this, we start with each gold 'daughter' replica separating from the 'mother' location (for example, X) and going to a different location (for example, Y). After sometime, and of its own volition, the 'daughter' returns back to the 'mother' location (for example, X). There were situations when two 'daughters' from different 'mothers' overlapped and were in significant physical contact with each other. It is important to note that they still remembered their 'mothers' and always returned back to them.

First, the process of physically disconnecting from the 'mother' structure and moving from location 'X' to 'Y', and then returning to 'X' without being subjected to any additional applied external force/stimulus indicated an apparent spatial memory register in each 'daughter' that is correlated with its correspondent 'mother'. Secondly, it implies an apparent auto programmed

"instinct" in these structures to **act** even in the absence of any internal force (i.e. non-living systems) or intentionally applied external force/stimulus. Thirdly, the local transport of these 'daughters' over space across the sapphire surface and precise (e.g. orientation) reassembly onto their original 'mothers' imply a strong correlation of their properties with respect to each other as well as to their 'mothers'. Finally, the observation that all the 'daughters' "*act*" in a coordinated manner with each other (i.e. *in sync*) to complete the process of navigation and re-assembly onto their 'mothers' implied a strong correlation between them as well. The apparent temporal and spatial synchronization demonstrated a dynamic communication and navigational feedback mechanism between each 'mother'/'daughter' pair. To re-iterate, the observed phenomenon requires the presence of a governing force field, is yet to be defined.

Regarding the experimental results for the nano-particles, the observation of a significant increase in their volume as a first response to an applied weak force is counter-intuitive. It is highly unlikely that the synchronized, time dependent appearance and disappearance of several significant physical features on the 'mother' structures in a uniform manner is governed by thermodynamic equilibrium considerations. Nano-particles transforming themselves into uncanny 3-dimensional "robot-like" forms as a force response, followed by their remarkable phenomenological display of the "oscillatory dance" between them tends to challenge the imagination. The inanimate "robot-like" structures which are replica copies of their complex forms seem almost incredulous. Adhering to the disciplined scientific method and staying cognizant of the Occam's razor principle, these phenomena seem inscrutable from the framework of currently established physical and thermodynamic principles.

**Conclusions:**

We report the preparation of 2-dimensional gold atomic sheets based nano-structures for the first time. These inanimate gold nano-ring structures are the first inorganic systems to exhibit an apparent *'homing instinct'* and an *'action'* based on that *'instinct'*, under the influence of a special but undefined force. We also report a remarkable oscillatory shape transformation phenomenon between the 3-dimensional 'robot-like' forms and their 'mothers'. High resolution imaging confirmed that 3-dimensional 'robot-like' structures are identical *replica copies of their highly complex forms.*

Besides the philosophical implications due to their, what appears to be, a higher level of *intelligence* exhibited by these structures, these findings lend validity to the new concept of "robots on chip". This would have the potential to fundamentally disrupt several technology fields. The sensitivity of these nano-structures to vibrations and acoustic waves, would enable remote dispatching of programmed "daughters' to target locations on the chip. The phenomenological observations reported in this manuscript are expected to stimulate significant fundamental research.

**Materials and Methods**

*A. Materials:*

Precursor material, namely, gold wires were purchased from Sigma-Aldrich (99.99%, highest available purity grade). Polished sapphire substrates (0.33 mm thickness, Single crystal, DSP Prime C-plane <0001> off M-plane <1-100> $0.2 \pm 0.1°$), were purchased from University Wafer Inc. USA.

B.  *Deposition of Gold Films***:**

Thermal evaporation process was used to deposited the gold thin films from a high purity gold wire onto the sapphire substrates.  First, these substrates were ultrasonically cleaned with isopropyl alcohol (IPA) and dried with a nitrogen-spraying gun. The substrates were then loaded into a plasma cleaning system (PDC-002, Expanded Plasma Cleaner, Harrick Plasma, USA) for a duration of 5 minutes.  Freshly cleaned substrates were transferred to the thermal evaporation chamber of Denton Vacuum LLC 502 B system. The base pressure of main chamber was ~ 3x $10^{-7}$ torr. Prior to the evaporation process, a lower current of ~50A was applied for 3 minutes for preconditioning the tungsten boat.   Film deposition rate of 1Å/sec at 72 A was calibrated by an in-built quartz crystal monitor (QCM) and maintained constant throughout the process. Gold coatings with different (QCM, Quartz Crystal Monitor) thicknesses (e.g. 10 nm, 25 nm, 30 nm, 50 nm (66 nm by X-ray reflectivity), 100 nm, 200 nm) on single crystal sapphire substrate were deposited. Subsequently, the deposited films were diced into 0.5 mm x 0.5 mm square shaped samples and post-deposition heat treated in a preheated oven (Carbolite 1200, UK). Post-deposition heat-treatment temperature ranged from 350 $^0$C – 600 $^0$C in air. The times of heat treatment were, 2s, 5s, 10s, 30s, 60s, 300s and 600s, respectively.

C.  *AFM characterizations of Gold Films:*

Atomic force microscopy (Park NX10, Park Systems Korea) was used in true non-contact mode for morphological characterization. This unique scan mode prevents potentially invasive tip-sample interaction during a scan. AFM scans were collected using Park SmartScan$^{TM}$ software

(Park Systems Korea) using ultra-sensitive (Super Sharp Standard NCH cantilevers, High Density Carbon tip, Manufacturer: NanoWorld AG, Switzerland) with a typical tip height of 10-15 μm and 2 nm tip radius, under ambient conditions. The physical dimensions of these cantilevers were 125 μm (length), 30 μm (width) and 4 μm (thickness), respectively. The force constant value and resonant frequency were 42Nm$^{-1}$ and 320 kHz, respectively.

In addition to conventional AFM imaging, Park AFM has in built capabilities to precisely measure the nano-scale mechanical properties of materials, namely, nano-indentation technique (load-displacement (P-H) curve) and an additional capability referred to as the Force vs Distance (F/D) Spectroscopy. F/D spectroscopy measures the vertical force applied by the tip to the surface during contact-AFM imaging, through the deflection of the cantilever as a function of the extension of the piezoelectric scanner. Typically, this technique employs a significantly smaller force than that used in conventional AFM nano-indentation experiments. In F/D spectroscopy, the dependence of the cantilever deflection on the extension of the piezoelectric scanner is directly correlated to the tip-sample interaction forces reflecting a surface mechanical property. In principle, the technique could be used to measure local variations in the elastic properties of the surface.

*D. TEM sample preparation and imaging:*

A facile route was implemented for the direct transfer of the gold nano-structures from sapphire substrate to the selected target for HRTEM imaging, the Quantifoil with 1.2μm holey carbon film coated Cu grids. This direct route of transferring the film ensured contamination free transfer. During the direct transfer, the support to the gold film is provided by TEM grid's carbon film. To bond the gold and carbon of the TEM grid, the grid is placed on top of gold and a drop of isopropanol (IPA) is gently placed on top of the grid ensuring that both the grid's carbon film and the underlying gold film is wet. The surface tension draws the gold and carbon together into close contact as IPA evaporated [28]. Residual IPA was evaporated at ambient conditions over a period of 24 hours.

Characterization of transferred gold film on the TEM grids was performed using a Talos F200X FEG Transmission Electron Microscope with a lattice-fringe resolution of 0.14 nm at an accelerating voltage of 200 kV equipped with CETA 16M camera. High resolution images of periodic structures were analysed using TIA software.

**Acknowledgments:** We acknowledge the valuable support provided by the Core Technology Platforms at NYU Abu Dhabi for the use of the instruments.

**Funding:** Funding for this work was provided by NYU Abu Dhabi.

**Figures**

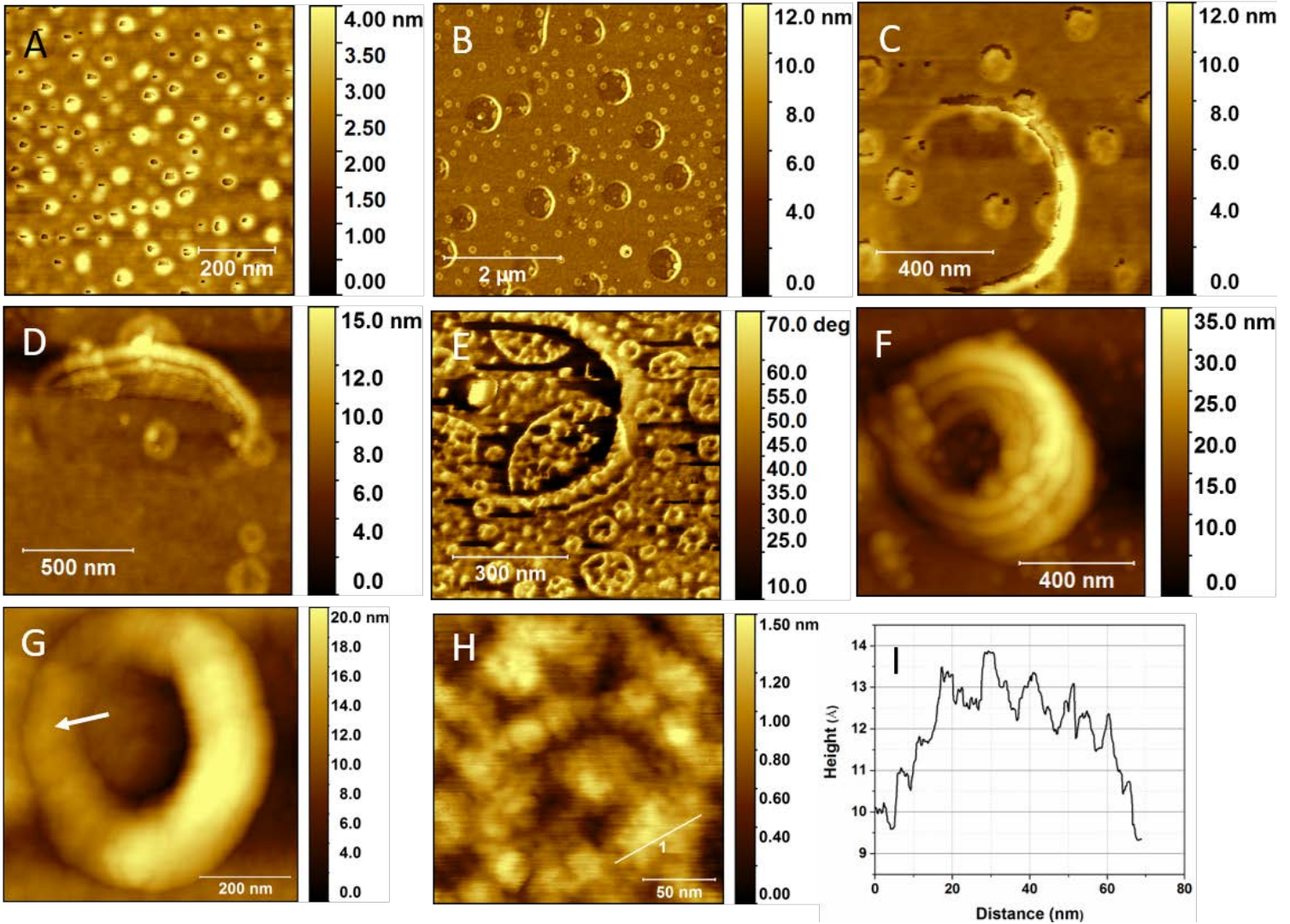

**Figure 1: Evolution of nano-ring structure**. Figure 1A shows several open dome like structures, with their edges pinned to the sapphire substrate. The "open-dome" like structures coalesce first to form larger, arc-like structures (Figure 1B and C). In Figure 1D, we are able to see evolution of the vertical growth stacks of these arc-like structures which are seen in more detail in AFM phase image (1E). In Figure 1F, we show the topographic image of an almost completed nano-ring structure, made up of stacks of several thinner rings. Figure 1G is the image of a completed nano-ring. High-resolution topographic AFM image of a selected area of the nano-ring in 1G (white arrow) is shown in Figure 1H. In Figure 1I, we measured the height profile across these smaller disc (building blocks) to be approximately 1.7 angstrom, implying that these discs are made up of one atomic layer of gold.

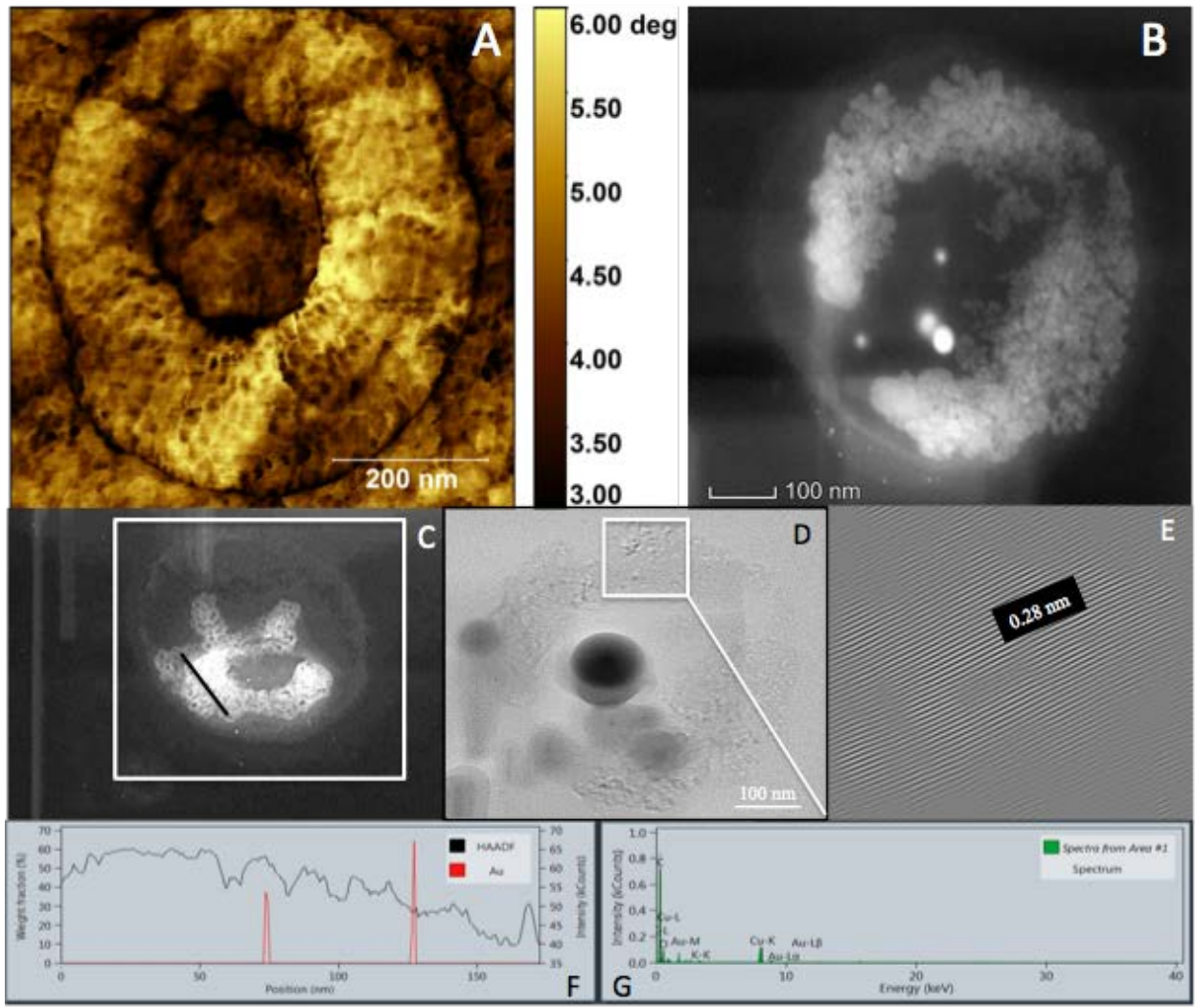

**Figure 2: Confirmation of gold nano-ring structures**. Figure 2A is an AFM phase image of a typical nano-ring structure. Figure 2B and 2C show the STEM images of two different nano-rings. Figure 2D shows the high-resolution surface imaging in bright field mode of the ring. The incident beam direction was normal to the specimen surface. Figure 2E shows the lattice parameters of the HRTEM image, reconstructed using IFFT. Elemental X-ray spectral mapping (EDAX) of Au shown in Figure 2F and 2G, was used to confirm the location of Au in the ring shown in Figure 2C (black line) confirming the presence of Au across the ring. Figure 2G shows the integration of characteristic peaks of Au centered at 2.1 and 9.7 keV in the EDS spectra.

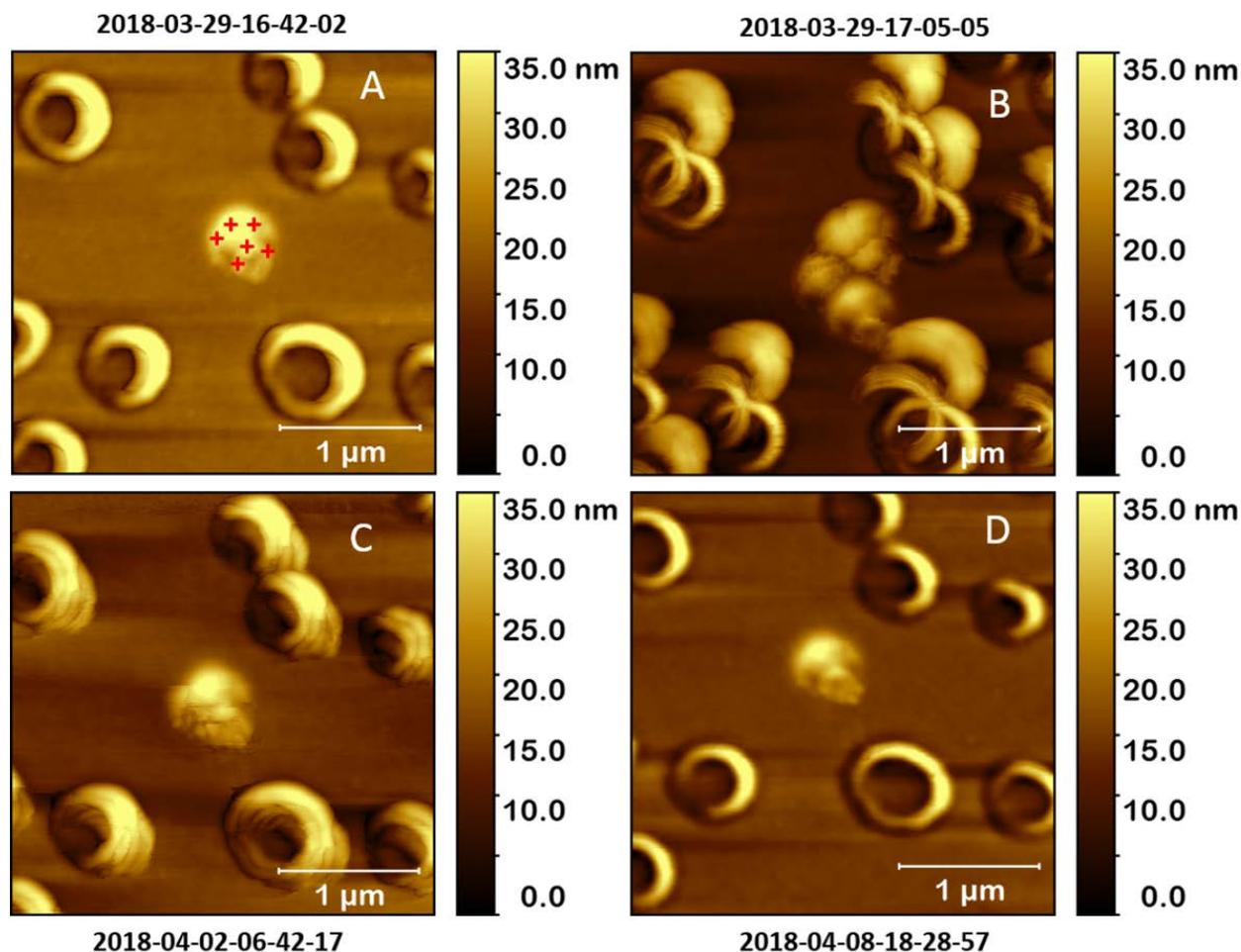

**Figure 3: Demonstration of "*homing Instinct*".** Figure 3A shows an AFM topographic image of the gold nano-ring structures, before being subjected to the weak force. Using the F/D spectroscopy technique, a force of 100 nN was applied to the "bowl-like" gold structure in the middle of the image frame marked by the red ink **(+)** marks in Figure 3A. The time and date of experiment is marked in the figure. In Figure 3B, we show the same image field, immediately after the application of the force. It is clear that application of the force generated several 'daughter' replicas from each 'mother' structure. Over a period of time, these 'daughter' replicas migrated back to the original 'mother'

structure location and assembled to re-create the original shape. Figure 3C shows the process of re-assembly in the intermediate stage. In Figure 3D, all the 'daughter' replicas have completed their migration back to their 'mother' structure and re-assembled to create the original shape. The final image 3D is remarkably identical to that of the starting image 3A.

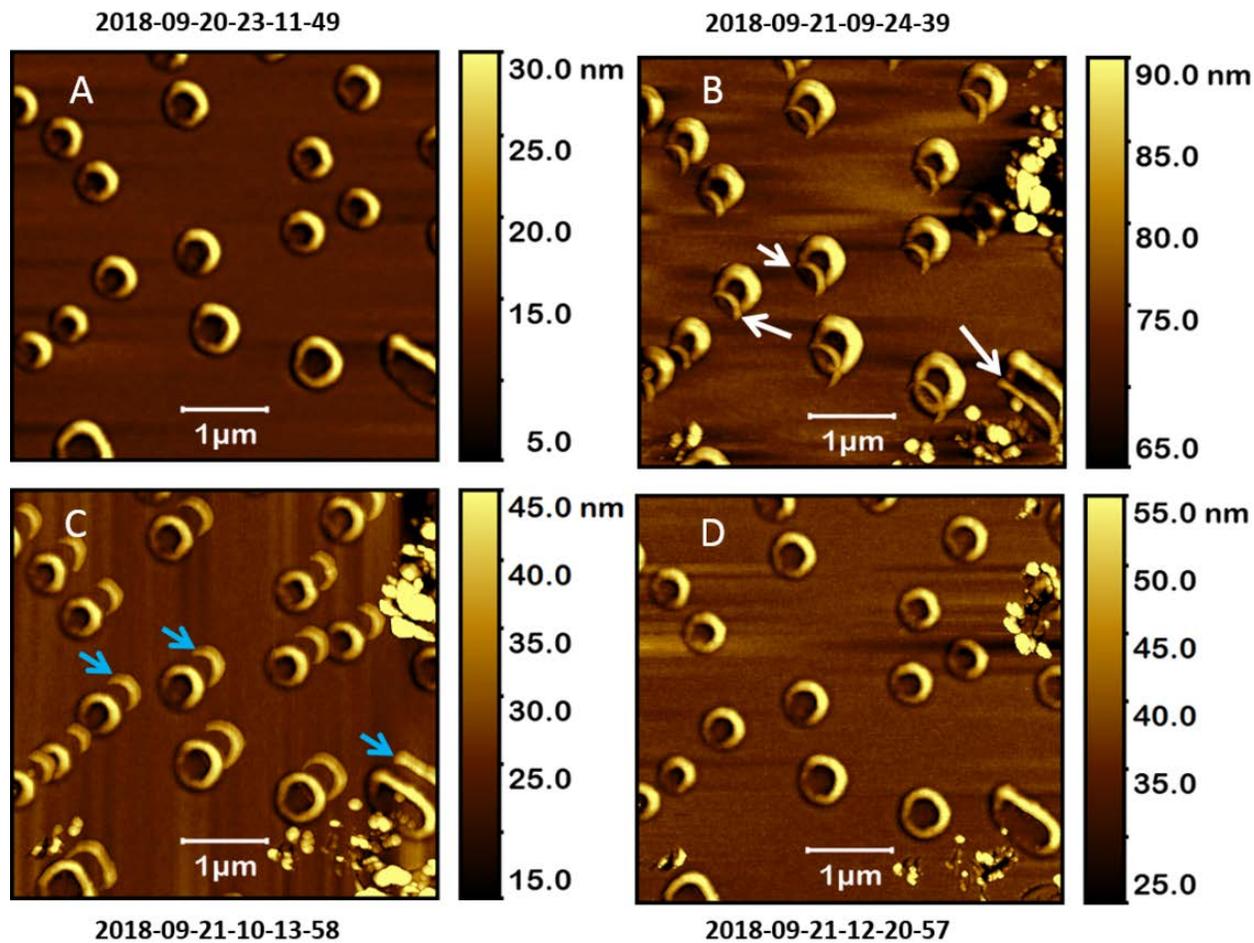

**Figure 4: 'Apparent memory register' and coherence.** Effect of exposure of nano-ring structures to 10 kHz acoustic wave. **A** shows nano-ring structures prior to acoustic wave exposure and **B** shows the image immediately afterwards. **C** is the image after 49 minutes and **D** is after 126 minutes. The colored arrows show the 'daughter' replicas. The observations confirm the presence of strong coherence between the nanostructures.

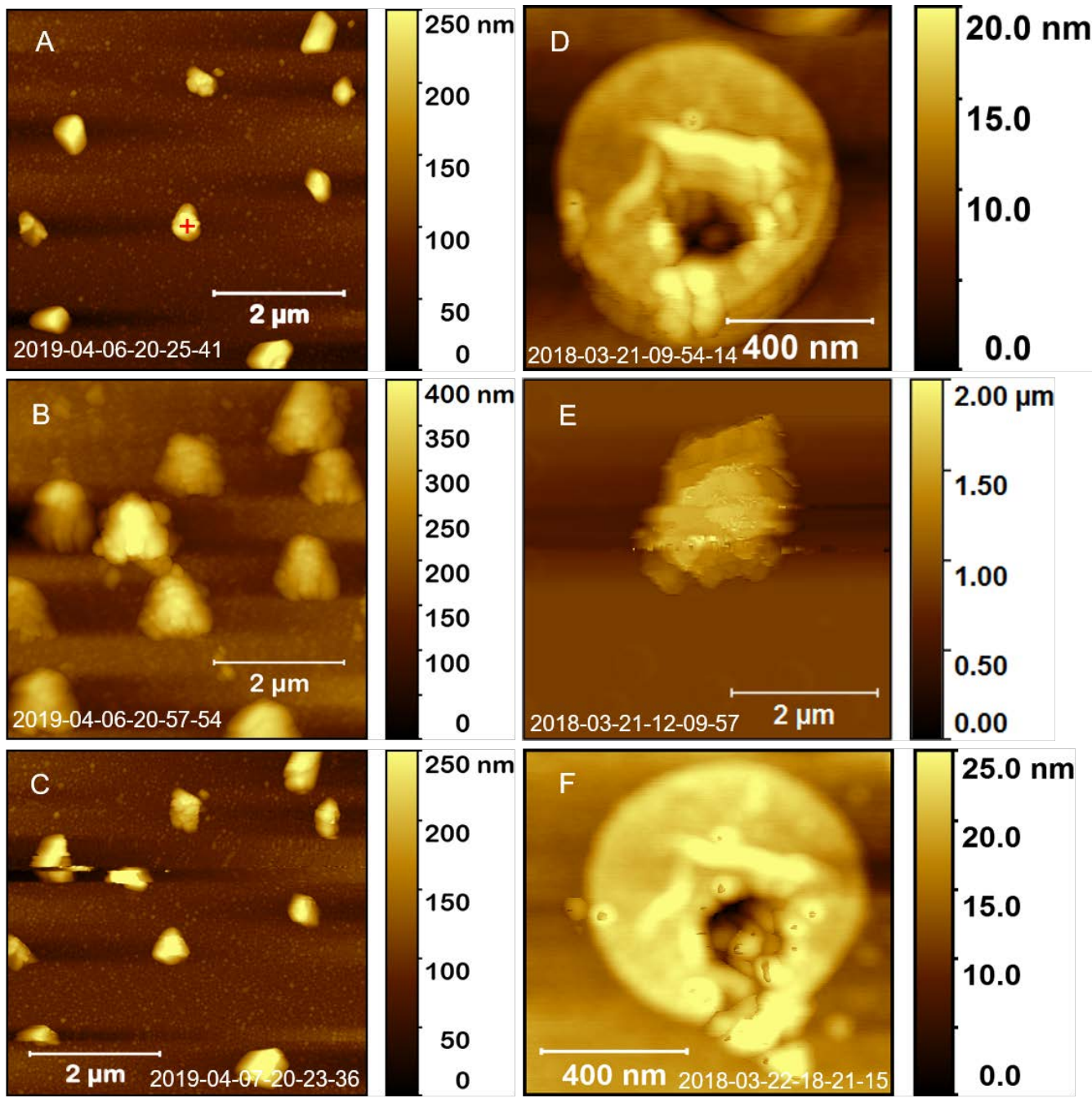

**Figure 5: Shape transformation phenomenon**. Demonstration of the change in shape/volume due to application of the force. In A we show an image of the nano-particles prior to the application of the force indicated by the red ink mark (+). B is the image of the same location immediately after the application of the force. It shows the transformation of each particle into a new form that is significantly larger in volume than that shown in A. C is an image of the same location as A and B showing the transformation of the structures in B returning back to the original structures in A. D

is an AFM image of a circular "hockey puck" shaped structure prior to application of the force. E is the image of the same location after the application of the force and F is the image of the location after 30 hours.

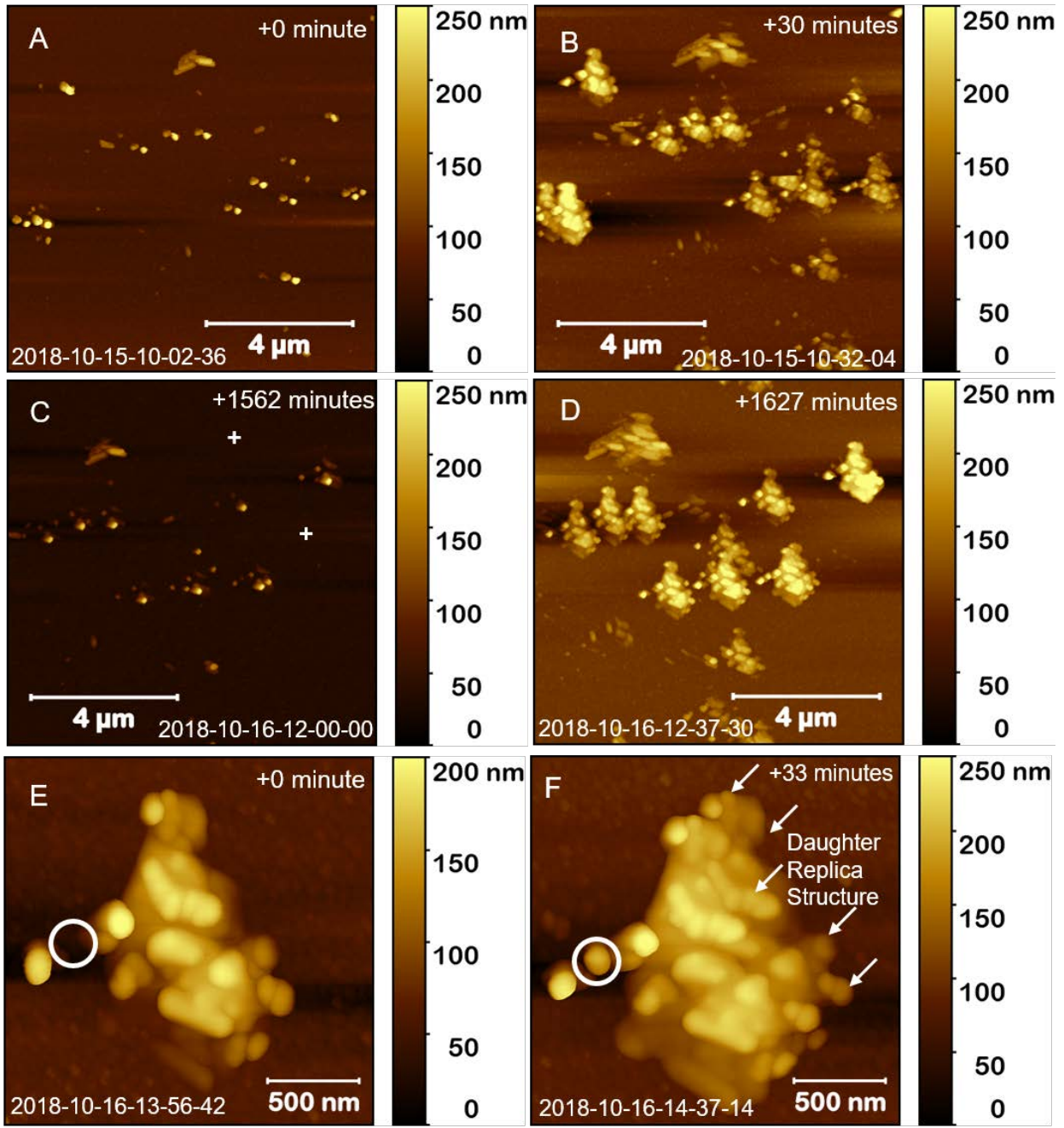

**Figure 6: Oscillatory shape transformation phenomenon**. In A we show the nano-particles prior to application of a force of 100 nN at a location marked by a white (+) ink mark. In B we show the "robot-like" structures due to application of the force. In C, we show the transformation of the "robot-like" structures into nano-particles. In D we show the transformation of the nano-particles in C into "robot-like" structures again. This data (A,B,C and D) are part of a larger data set of images collected at different time intervals, over a period of 26 hours. They clearly demonstrated

the unique phenomenon of "oscillatory shape transformation". Higher magnification images of a complex 3-dimensional "robot-like" structure is shown in E and F. F is a re-scan of the same image shown in E after 33 minutes. It is evident that the structure in E has replicated a copy of itself.

# Supplementary Materials:

## Apparent Spatial Memory Register and Oscillating Shape Transformations in Gold Nanostructures

Sudhir Kumar Sharma,[1] Renu Pasricha,[2] James Weston,[2] Thomas Blanton[3] and Ramesh Jagannathan[1*]

.
Correspondence to: rj31@nyu.edu

**This file includes:**

**Research Summary**

We report our experimental findings for gold nano-ring and nano-particles.

**Nano-rings:**

We repeatedly observed that, after being subjected to a momentary weak force, nano-ring 'mother' structures undergo the process of creating 'daughter' structures that are replicas of their respective 'mothers'. Over time, the 'daughters', without fail, navigated back to their original 'mothers' and re-created the original 'mother' structure. That is, each gold 'daughter' replica separates from the 'mother' (for example, location X) and goes to a different location (for example, Y) and after some amount of time, and of its own volition, returns back to the 'mother' (for example, location X). Even if two 'daughter' replicas from different 'mothers' overlap and are in significant physical contact with each other, they remember their 'mothers' and always return back to them.

In order to conceptually illustrate these experiments clearly, let us assume that we have four 'mother' structures, (e.g. A, B, C and D). For logical clarity, let us also assume that, by applying a momentary weak force on one 'mother', each 'mother' simultaneously generated four 'daughters' (for example, $A_i$, $B_i$, $C_i$ and $D_i$ where i = 1,2,3 and 4) that were physically scattered around and away from their original 'mother'. Over time, all the $A_i$ 'daughters' would return only to A and re-form the original structure in its original location. We would observe the same behavior for each of the other three structures as well. That is, the 'daughters' Bi, Ci and Di would navigate back to their respective 'mothers', B, C and D respectively. Even if a 'daughter' (e.g. Ai) was in significant physical contact with another 'daughter' (e.g. Bi), Ai and Bi would each return only to their respective 'mothers' A and B. Navigation of these 'daughters' from all four 'mothers' would also be synchronized with each other and their respective mother structures.

First, the process of physically disconnecting from the 'mother' structure and moving from location 'X' to 'Y', and then returning to 'X' without being subjected to any additional external force/stimuli indicates an apparent spatial memory register in each 'daughter' that is correlated with the correspondent memory register in its 'mother'. Secondly, it implies an apparent auto programmed "instinct" in these structures to **act** even in the absence of any internal force (i.e. non-living systems) or intentionally applied external force/stimuli. Thirdly, the local transport of these 'daughters' over space across the sapphire surface and precise (e.g. orientation) reassembly onto their original 'mothers' imply a strong correlation of their properties with respect to each other as well as to their 'mothers'. Finally, the observation that all the 'daughters' "*act*" in a coordinated manner with each other (i.e. *in sync*) to complete the process of navigation and re-assembly onto their original 'mothers', imply a strong correlation between them as well. The apparent temporal and spatial synchronization imply a dynamic communication and navigational feedback mechanism between each 'mother'/'daughter' pair and between the collective 'mother'/'daughter' pairs as well. To re-iterate, the observed phenomenon requires the presence of a governing force field that is yet to be defined.

Based on our experimental observations with nano-rings, we conclude the following: these inanimate gold structures are the first inorganic systems to exhibit an apparent *"homing instinct"* and an "*action*" based on that "*instinct*", under the influence of a special but undefined force. The ability of the 'daughter' replicas to "*act in sync*" with each other is an added level of sophistication to this "*programmed instinct"*.

**Potential Impact**: Besides the philosophical implications due to the higher level of *intelligence* exhibited by these inanimate gold structures, from a practical viewpoint, it would lend validity to the concept of "robots on chip", with potential to fundamentally disrupt several technology fields. For example, one could imagine having them embedded in the chip architecture and due to their sensitivity to vibrational and acoustic waves, communicate with them remotely and dispatch programmed "daughters' to target locations on the chip. They would accomplish specific tasks and then return back to their 'mother' location. The phenomenological observations reported in this manuscript should stimulate significant fundamental research in this new field.

**Nano-particles:**
Momentary application of a weak force on a gold nanoparticle resulted in the spontaneous transformation of each nanoparticle into a highly complex and "robot-like" (e.g. Dalek) larger three-dimensional form. All the "robot-like" structures had an uncanny resemblance to each other. Without the application of any additional external force or stimuli, we observed a remarkable phenomenological display of, for the lack of a better word, an oscillatory dance between these "robot-like" forms and their original "mother' nano-particles, that lasted for several days. These "robot-like" structures were also capable of creating replica copies of themselves. In general, the nano-particles responded to an applied force by creating various complex and novel 3-dimensional structures/forms that exhibited this phenomenon.

In many experiments, we found the first response of the nano-particles to applied pressure was to increase their volume significantly (e.g. 100X) which is counter-intuitive. In many cases, they would add additional features to their body with time and at some point, the whole process would reverse itself until the original form was recovered. It is highly unlikely that the synchronized, time dependent appearance and disappearance of several significant physical features on the 'mother' structures in a uniform manner is governed by thermodynamic equilibrium considerations. Nano-particles forming the uncanny "robot-like" forms as a force response, followed by the remarkable phenomenological display of the "oscillatory dance" between them challenges one's imagination. The inanimate "robot-like" structure creating a *temporal replica copy* of itself seems almost incredulous. Adhering to the disciplined scientific method and following the Occam's razor principle, we find these phenomena to be inscrutable from the perspective of current understanding of physical and thermodynamic principles.

**Fig. S1.**

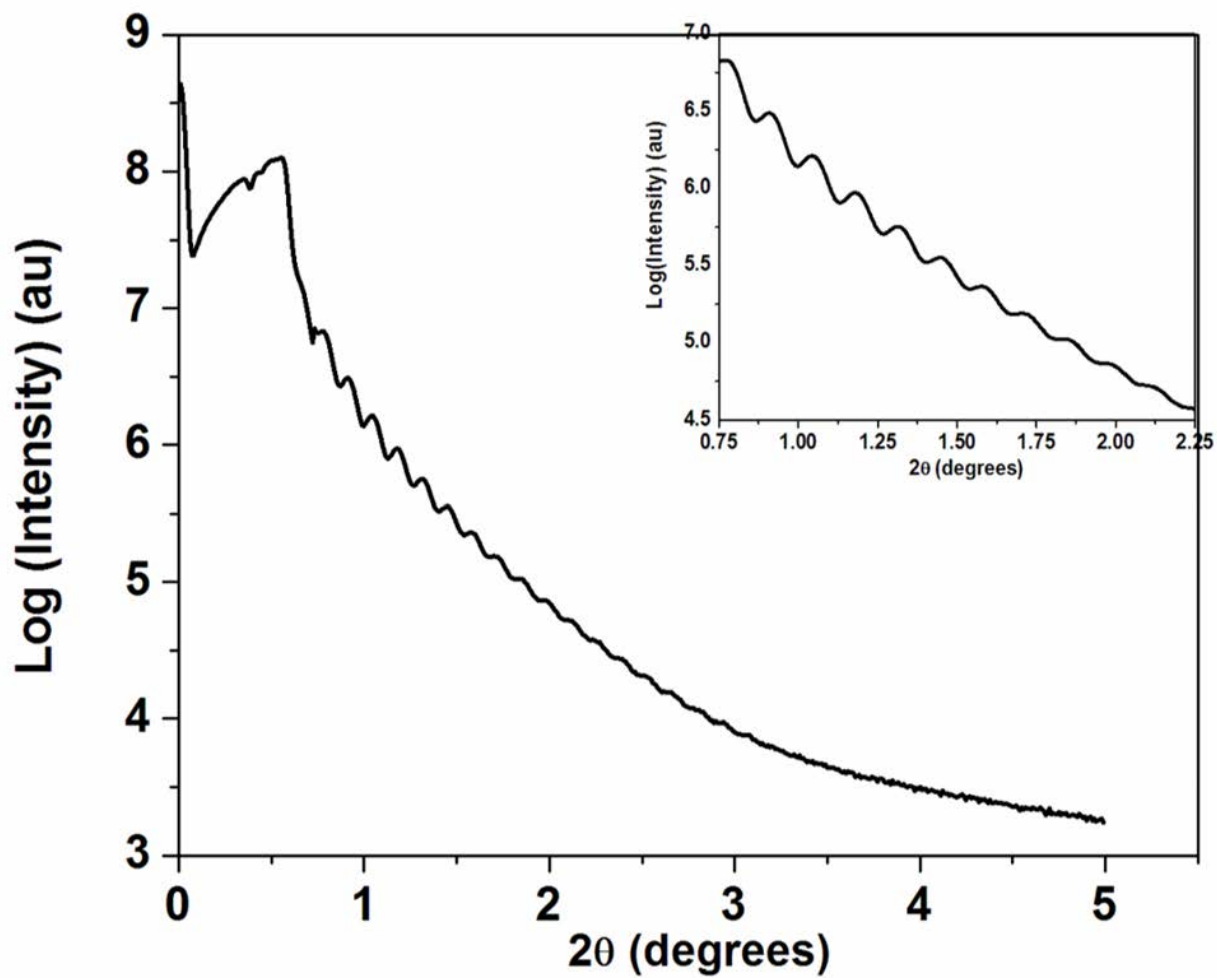

**Figure S1: X-ray reflectivity (XRR) pattern for as-deposited gold film on (0001) sapphire substrate:** Measured film thickness, 66 nm. Inset: selected range of XRR pattern. [Data collected at room temperature using a Bragg-Brentano diffractometer, Cu X-ray tube, reflection mode geometry]

**Fig. S2.**

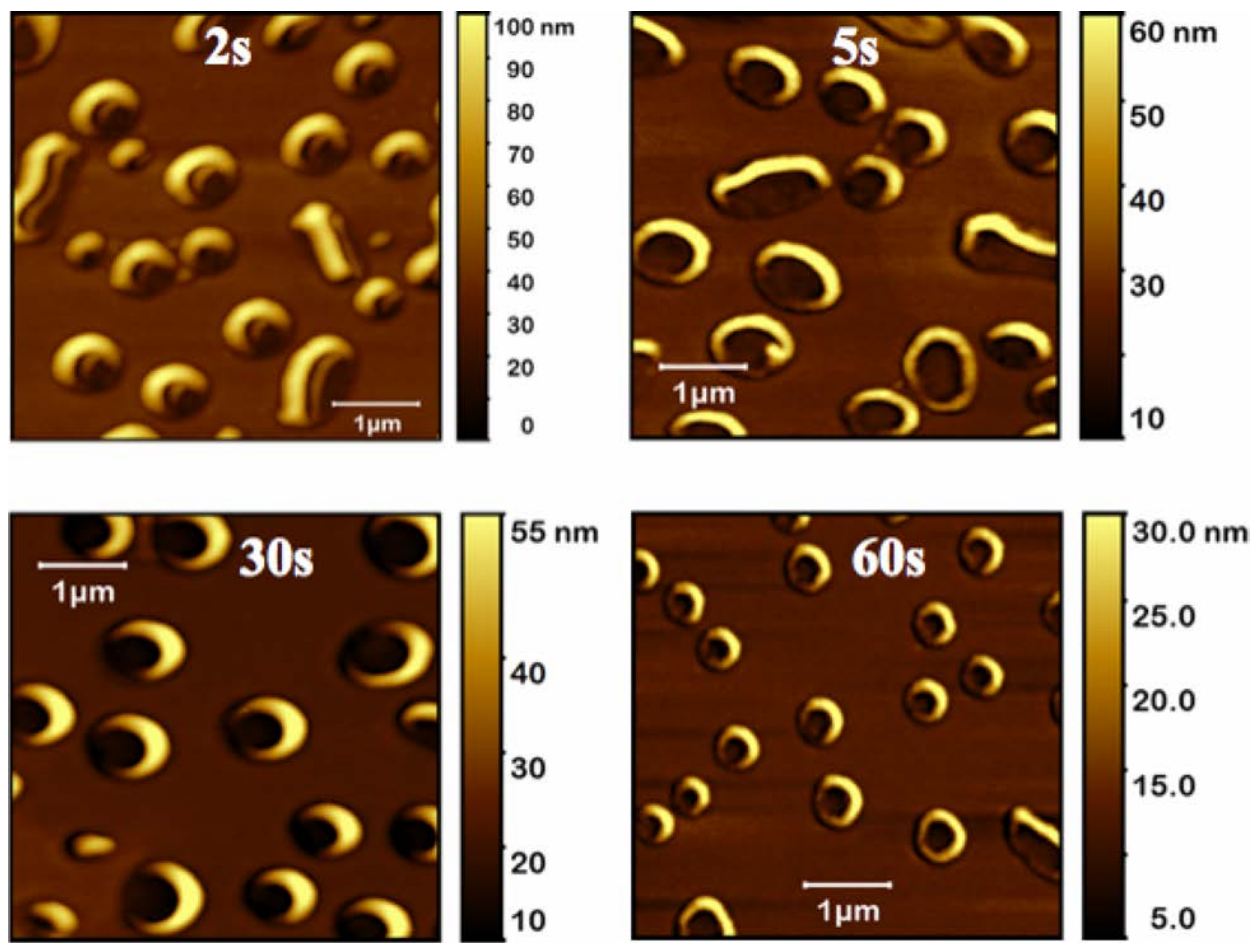

**Figure S2: Effect of time of 475$^0$C heat treatment of gold film:** Gold nano-ring structures were observed when the 66 nm gold film was treated to 475$^0$C for 2s, 5s, 30s, and 60s, respectively.

**Fig. S3.**

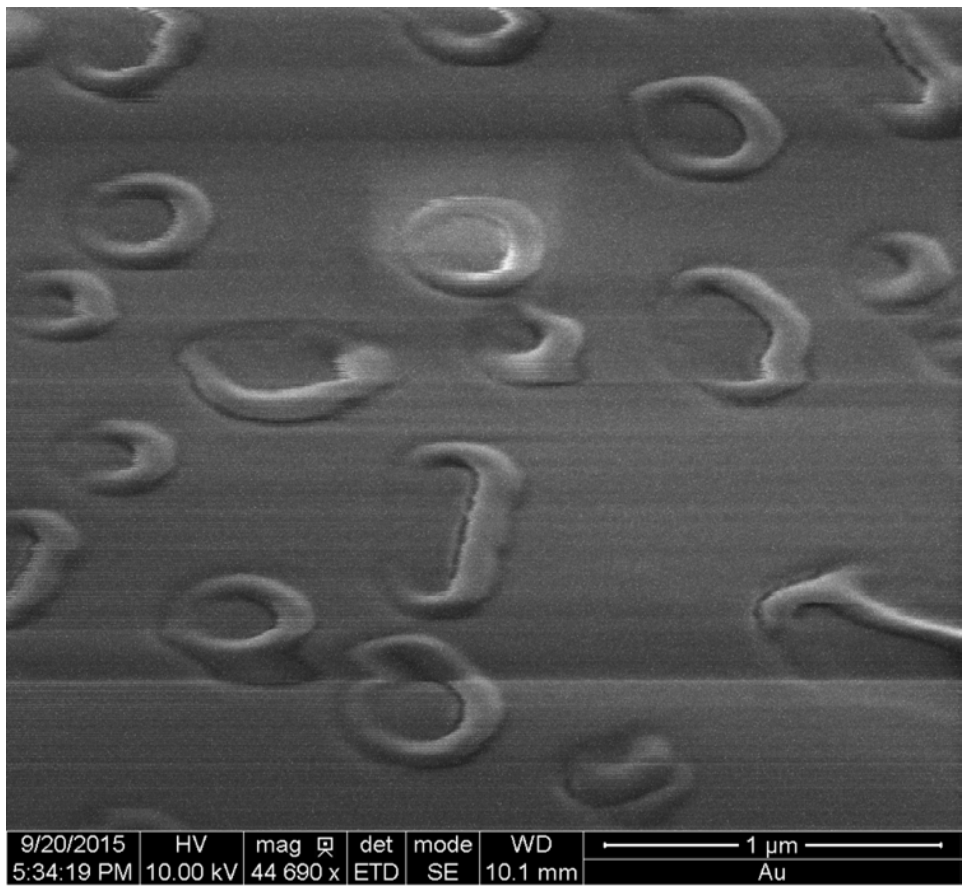

| 9/20/2015 | HV | mag ⊡ | det | mode | WD | 1 µm |
|---|---|---|---|---|---|---|
| 5:34:19 PM | 10.00 kV | 44 690 x | ETD | SE | 10.1 mm | Au |

**Figure S3: Effect of SEM imaging of on the gold nano-ring structures:** This figure shows the susceptibility of the gold nano-rings to e-beam exposure during SEM imaging.

**Fig. S4**

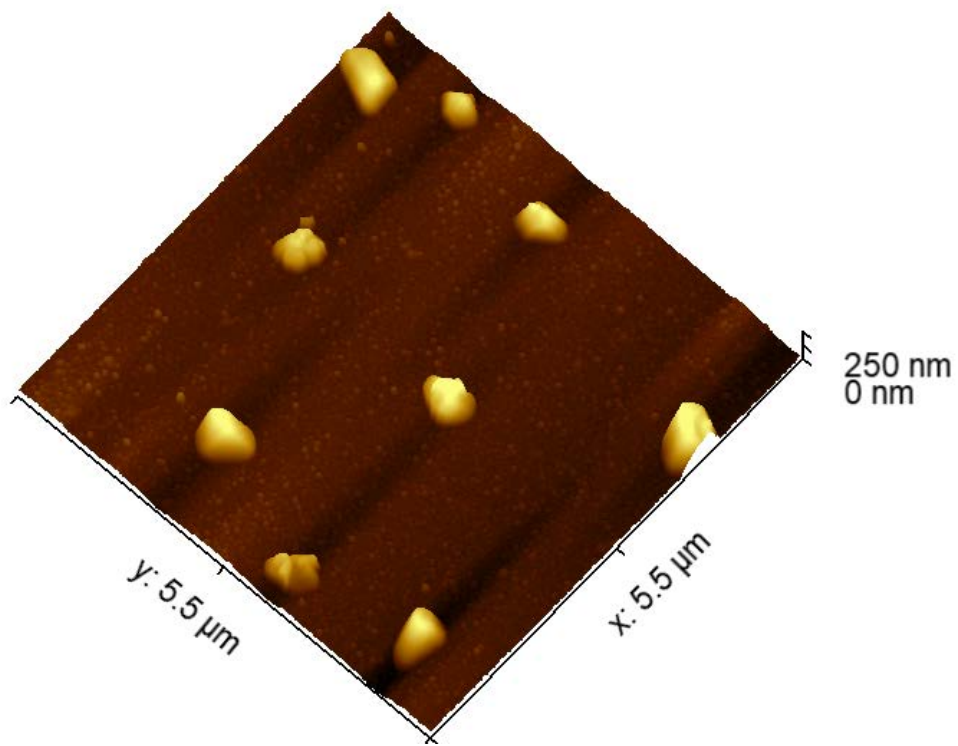

250 nm
0 nm

y: 5.5 μm

x: 5.5 μm

**Fig. S5.**

**Figure S4:  Nano-particles:** Sample AFM image of nano-particles observed when 200 nm (QCM) thick gold coatings were heat treated at 550 $^0$C for 30 seconds**.**

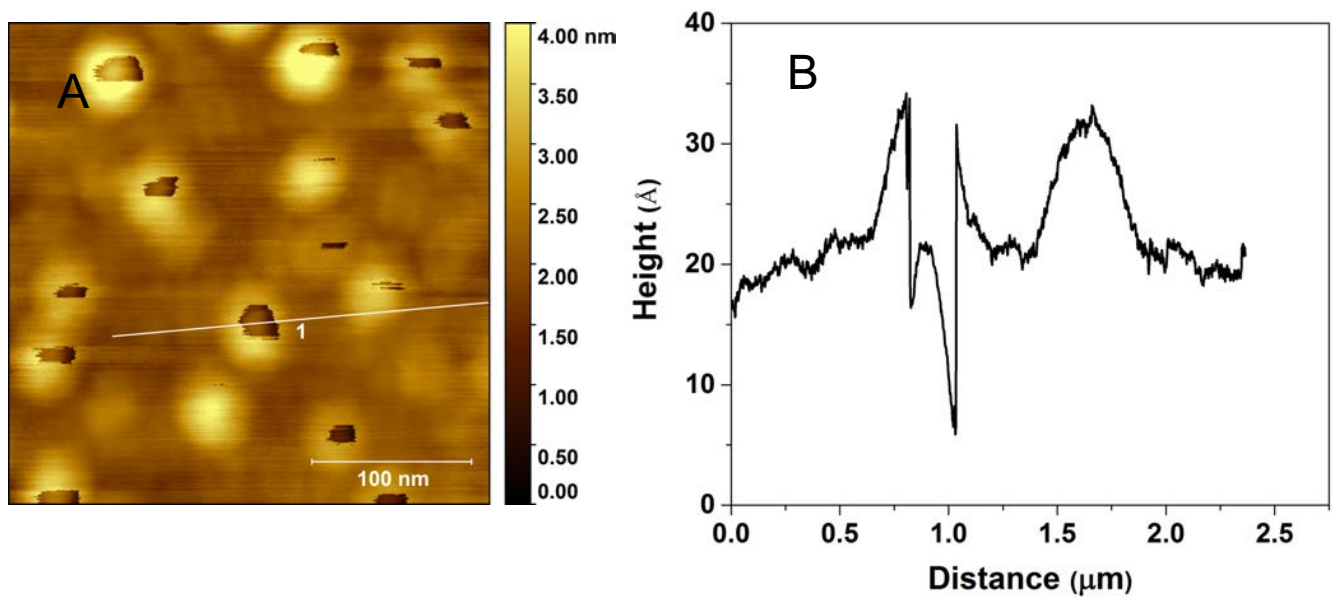

**Figure S5: Demonstration of the "open dome" structure:** S5A, is a higher magnification topographic image of the structures in Figure 1A. We measured the quantitative height profile across one of the open-dome like structures along line 1.  In Figure1B, we show the height profile along line.   The open-dome like structures rise from the pinned edges to a height of about 2 nm over an axial distance of 1μm on either side of the open top.

**Fig. S6:**

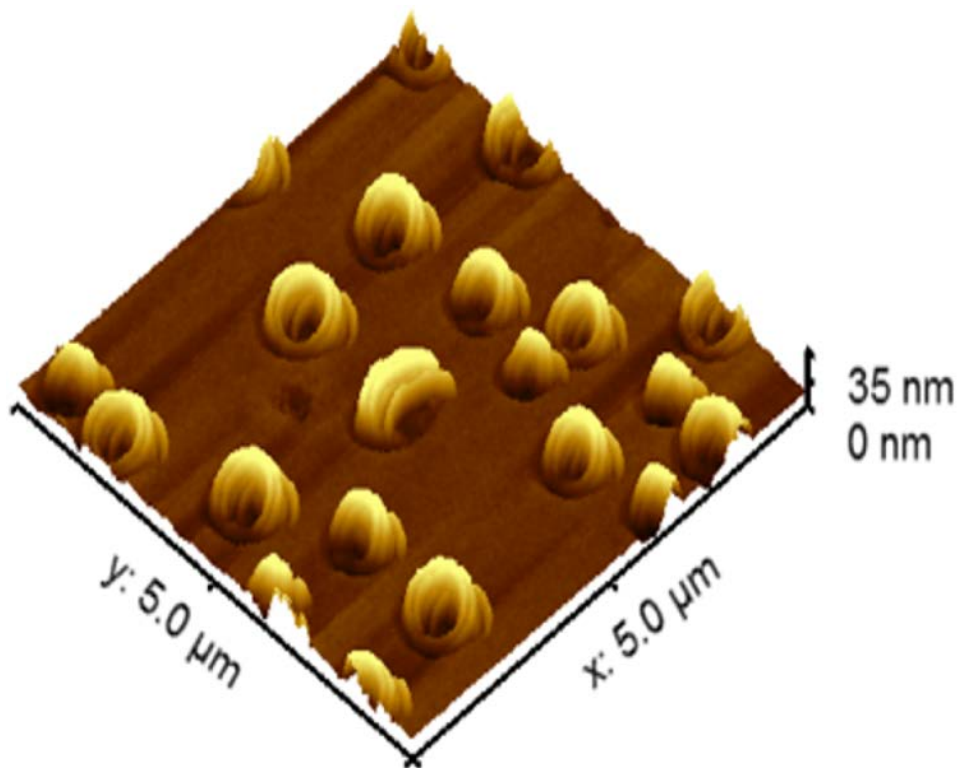

35 nm
0 nm

y: 5.0 μm    x: 5.0 μm

**Figure S6: 3-D representation of AFM images of the gold nano-rings:** 3-D view of the gold nano-ring structures demonstrate their asymmetric growth morphology. The nano-ring structure in the center of the image frame shows fine features of rapid growth, perpendicular to the substrate surface.

**Fig.S7**

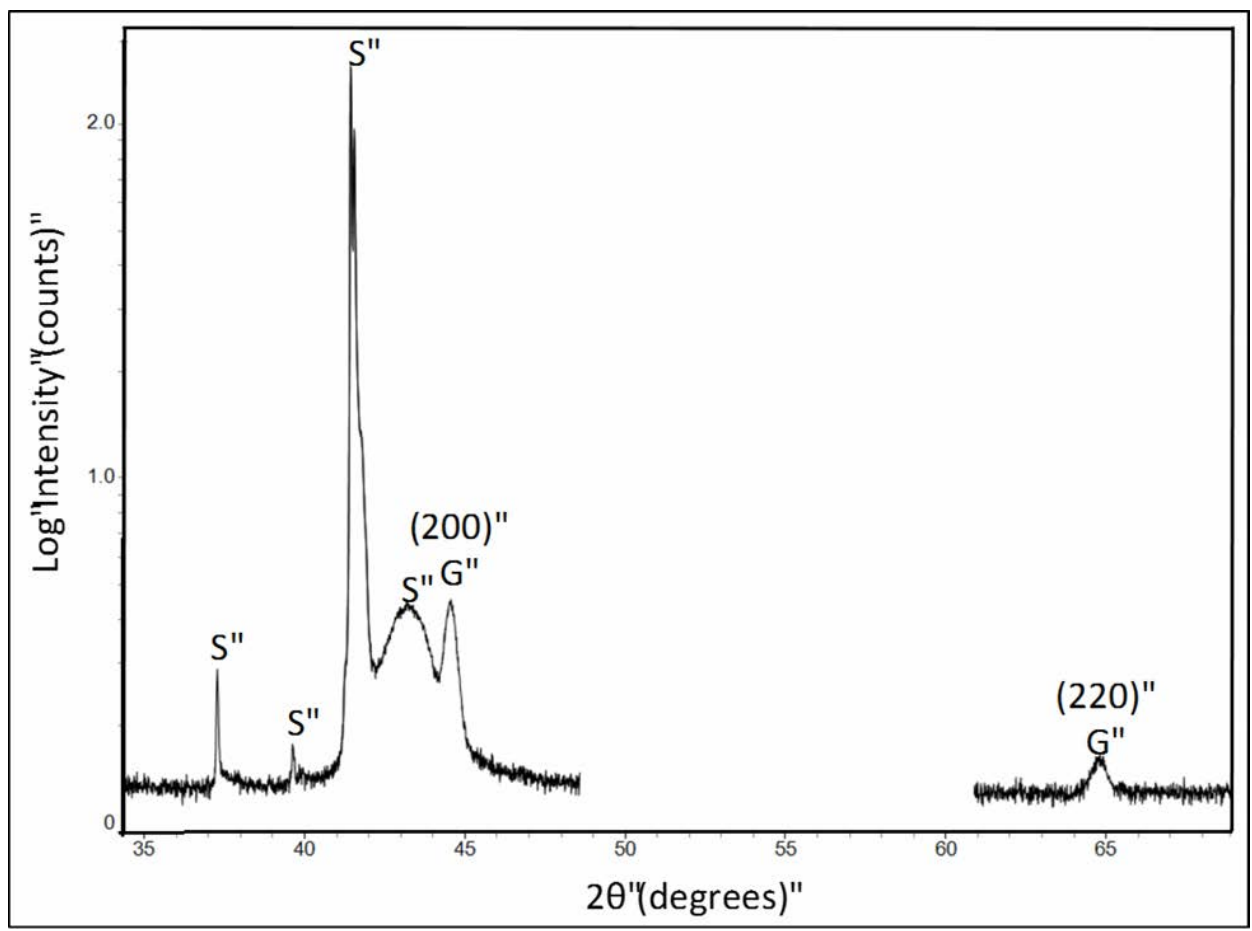

**Figure S7: Identification of the gold nano-rings by X-ray diffraction:** Selected range X-ray diffraction (XRD) patterns for gold on (0001) sapphire substrate after exposing sample to 475 $^{0}$C for 60s. Diffraction peaks marked S are due to the sapphire substrate. Diffraction peaks marked G are due to gold, with the (200) and (220) diffraction peaks identified. [Data collected at room temperature using a Bragg-Brentano diffractometer, Cu X-ray tube, reflection mode geometry]

**Fig. S8.**

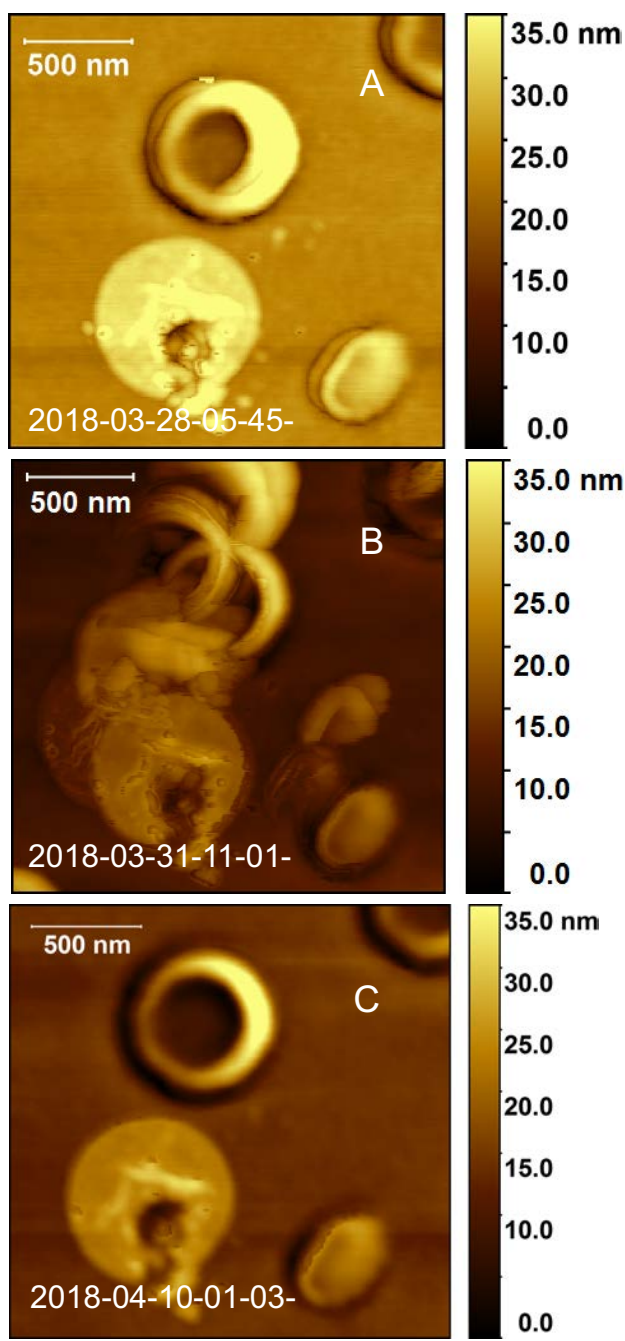

**Figure S8:  Unique spatial memory register and "homing instinct" of "daughter" structures:** S8A shows gold structures before the application of the weak force.  S8B is the image of the gold structures after the application of the force.  It is clear that the fragment replicas from two of the "mother" structures significantly overlap with each other. S8C shows the remarkable property of spatial memory and migration to their respective "mother" structures of the fragment replicas.  The fact that they are in significant contact and overlapping with each other did not prevent them from migrating back to their original "mother" structure.

**Fig. S9.**

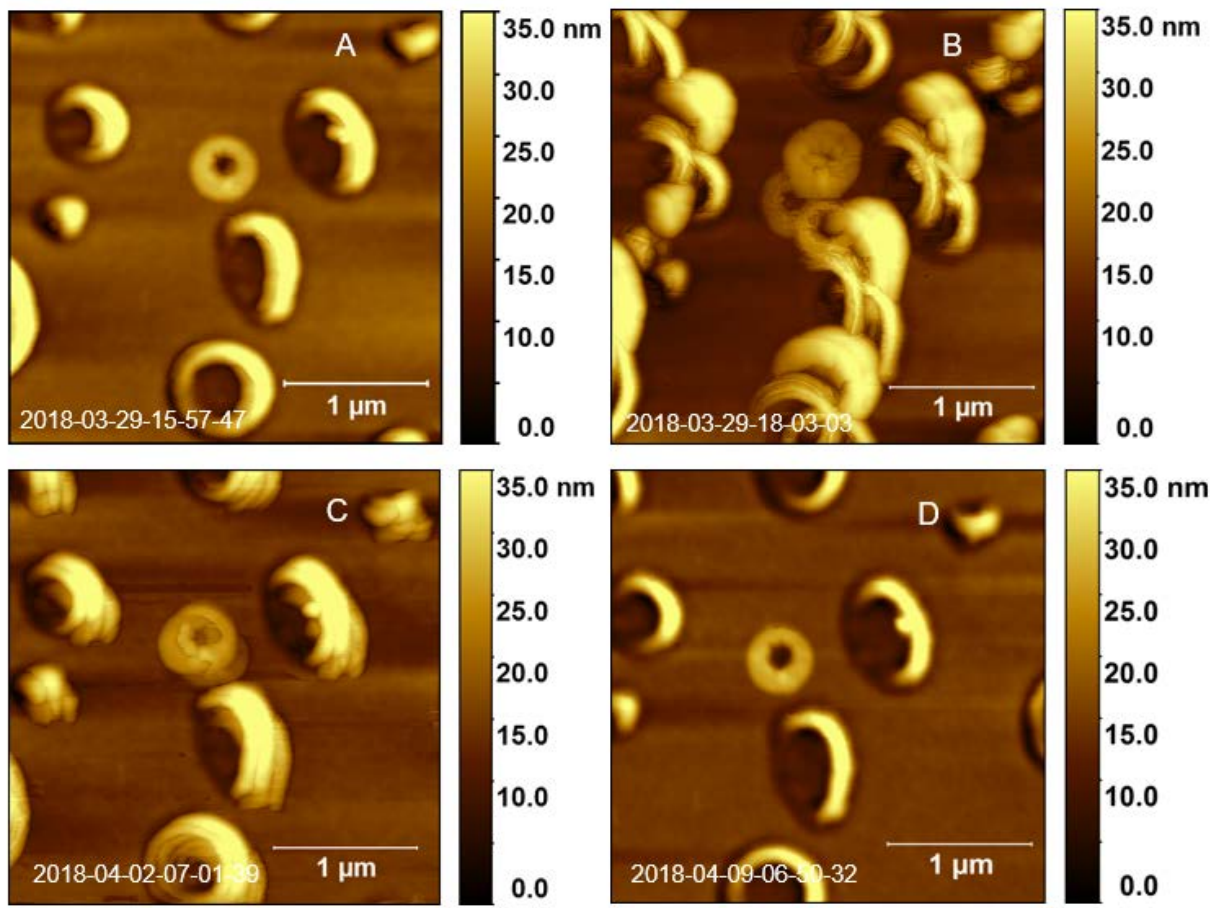

**Figure S9: Simultaneous observation of the "*homing Instinct*" at a location 45µm away from the point of application of the force, as marked in Figure 4A**: In S9A we show an AFM topographic image of the gold nano-ring structures, before the application of the force. In S9B, we show the same image field, immediately after the application of the force. It is clear that application of the force generated several 'daughter' replicas from each 'mother' structure, similar to the observation made from Figure 4B. Over a period of time, these 'daughter' replicas migrated back to the original 'mother' structure location and assembled to re-create the original shape, as was also observed in Figure 4C and 4D. S9C shows the process of re-assembly in the intermediate stage. In S9D, all the 'daughter' replicas have completed their migration back to their 'mother' structure and re-assembled to create the original shape. The final image S9D is remarkably identical to that of the starting image S9A.

**Fig. S10.**

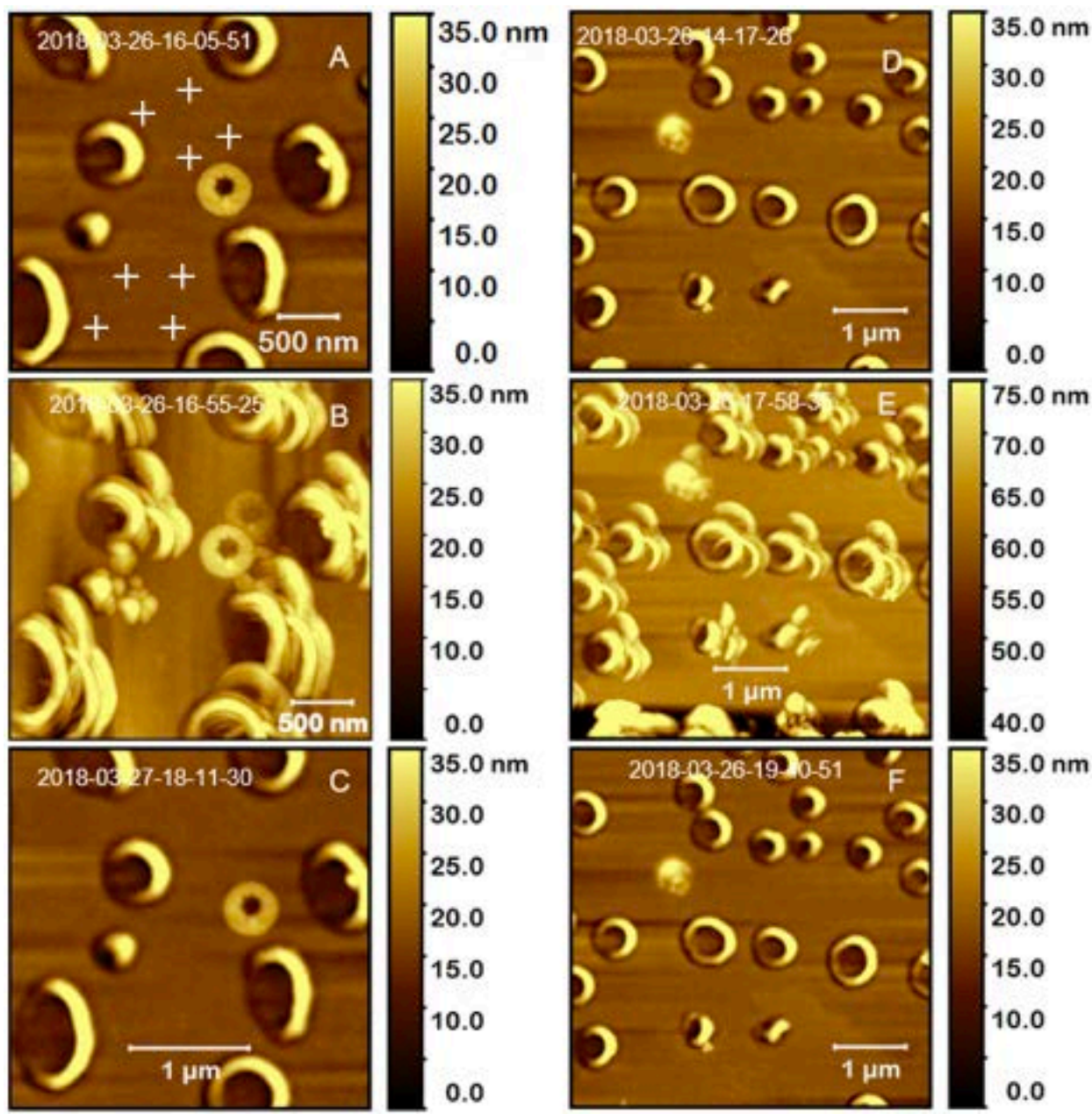

**Figure S10: Observation of the identical "mother"/daughter" phenomenon at a location spatially well removed from the point of application of the weak force:** **A** shows the nano-ring structures prior to the application of the weak force at locations marked by a white colored "+" sign . **B** shows the same nano-ring structures 50 minutes after the application of the weak force. **C** shows the same nano-ring structures after 2 hours. **D** shows the nano-ring structures that were 45μm away from the nano-structures shown in **A**, prior to the application of the weak force in the locations marked in **A**. **E** shows the nano-structures 1hour and 53minutes after the application of the force in **A**. **F** shows the nano-ring structures 3hours and 35minutes after the application of the force in **A**.

**Fig. S11.**

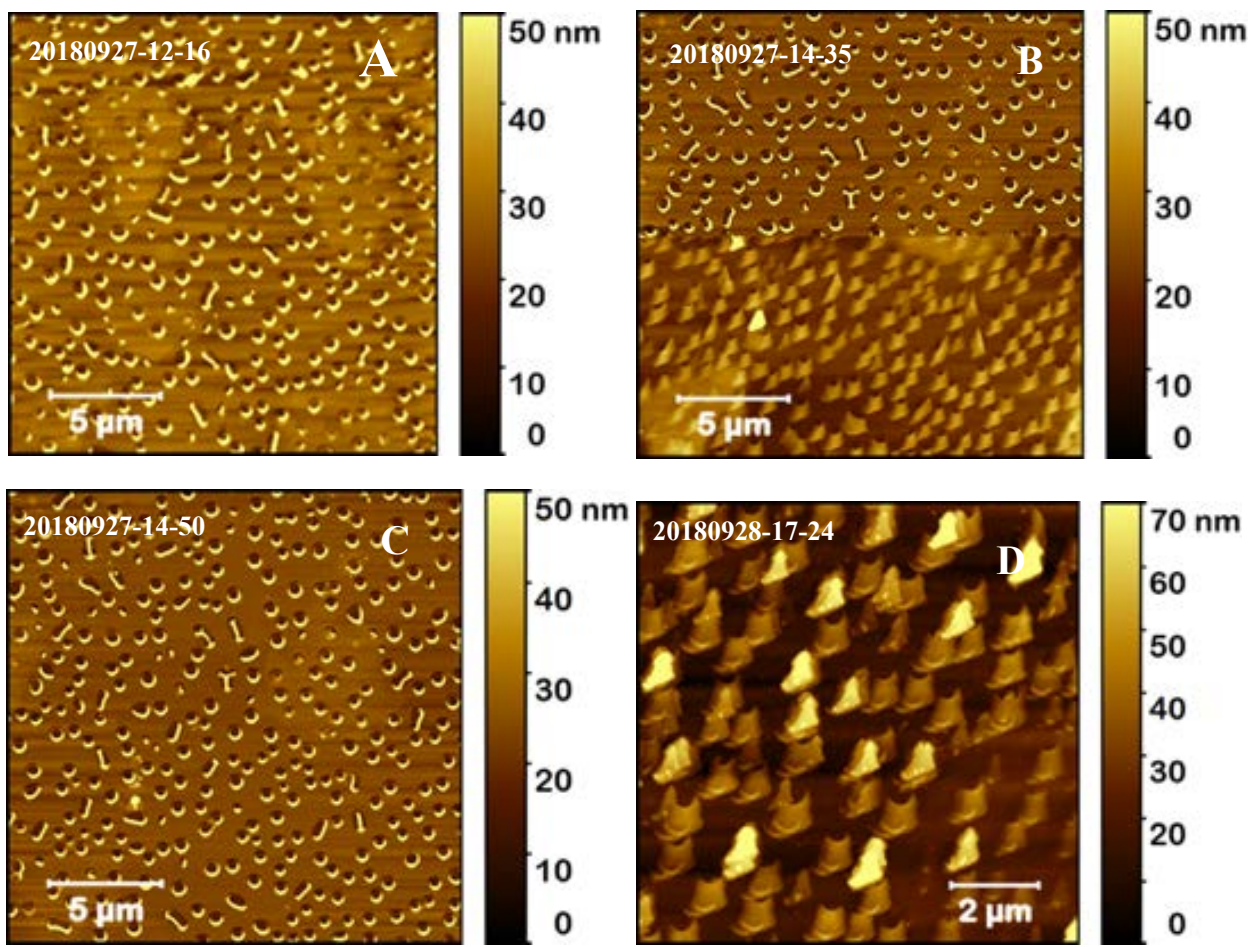

**Figure S11: Effect of exposure of the gold nano-rings to vibrations:** Figure S11A is an AFM image of the nano-rings prior to exposure to vibrational waves (125Hz,10s). Figure S11B is the image immediately after vibration exposure, showing the synchronized shape transformation during the middle of the scan. Figure S11C is the image from a subsequent scan confirming the synchronized transformation of all the truncated "cone-like" structures into nano-rings. Figure S11D shows the repeat transformation of the nano-rings back to the truncated "cone-like" structure.

**Fig. S12.**

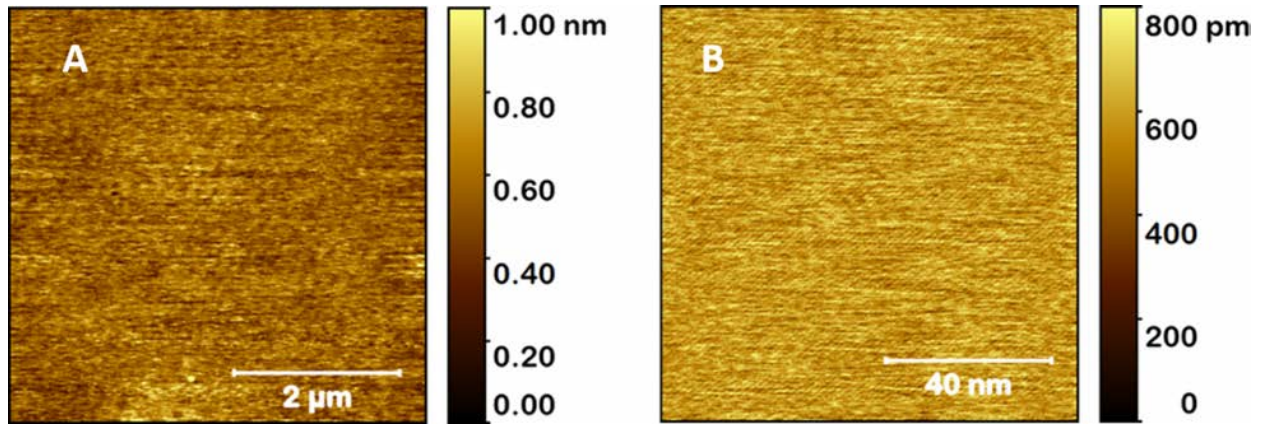

**Figure S12: AFM image of clean sapphire substrate:** A is a 5 μm scan and B is 100 nm scan. The average roughness values are found to be as follows: Ra=102 pm, Rrms=130 pm for the 100nm scan shown in B. The sapphire substrate used in our studies was always plasma cleaned prior to gold film deposition.

**Fig.S13:**

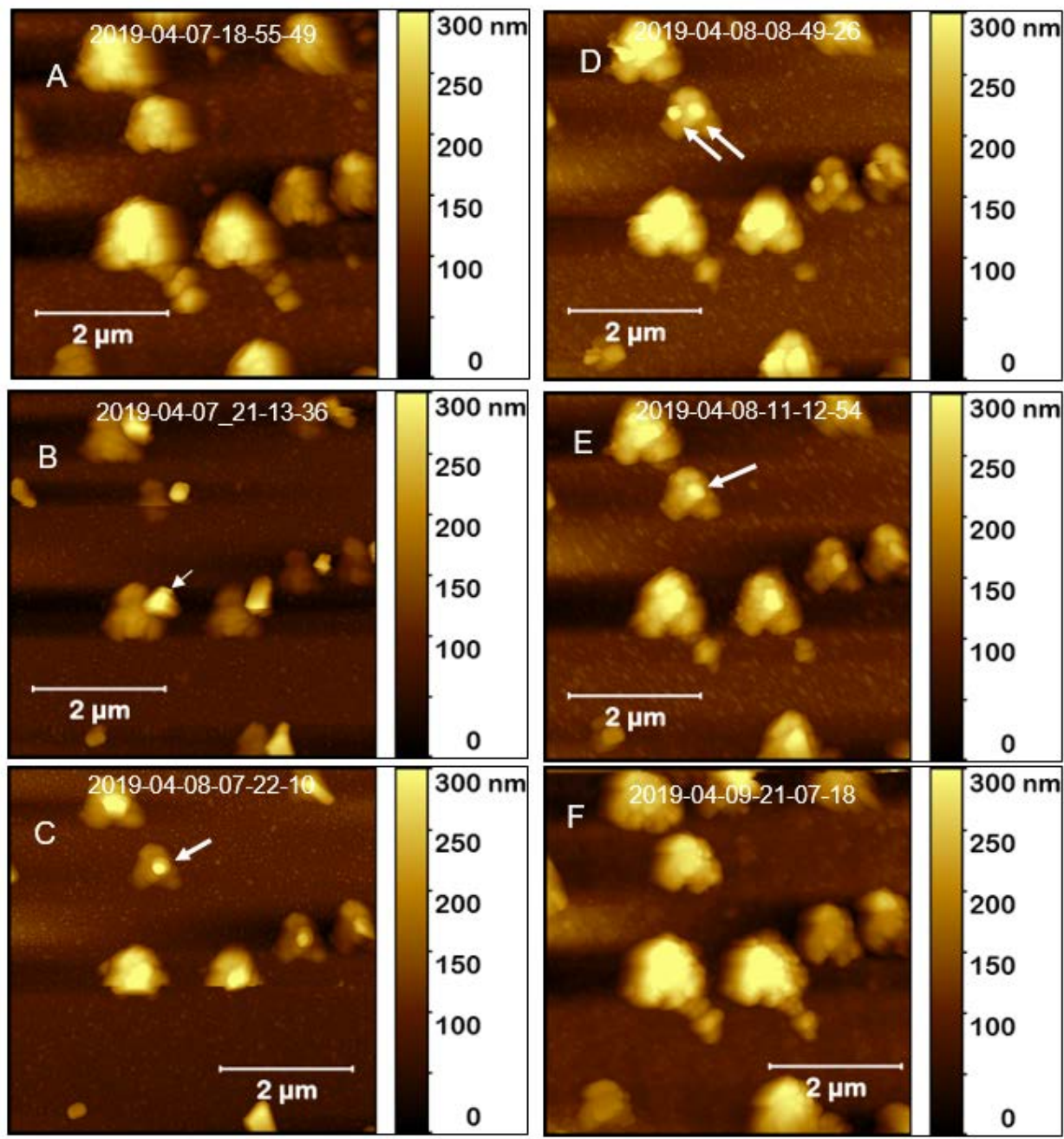

**Figure 13: Synchronized oscillatory transformation of 3-dimensional structures:** In A we show the AFM image of (mother) structures that resulted due to the application of a force to the nano-particles. B is the image of the same structures after a few hours showing each 'mother' with a significant 'daughter' particle attached to a corner. C is the image of the same location after a few hours. The 'daughter' particle has now moved to the center of each 'mother' structure. D is the image of the same location after a few hours showing each 'mother' with two 'daughter' structures. Similarly, imaging the same location after a few hours (E) showed each 'mother' with only one 'daughter' as in C. F is the image of the same location, after several hours showing the disappearance of the 'daughter' structures altogether and the 'mother' structures returning to their original state as in A.

**Fig. S14**.

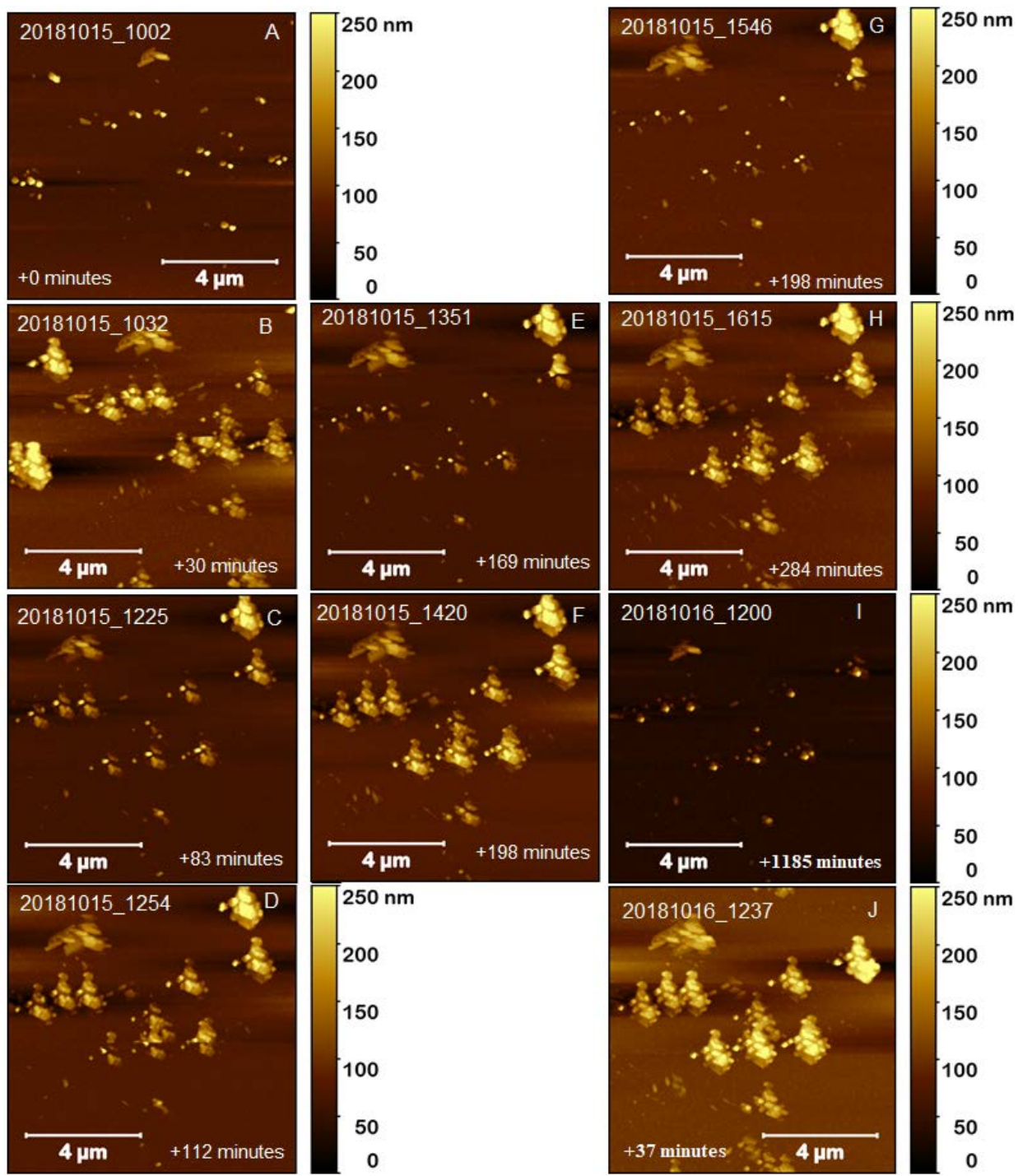

**Figure S14: Demonstration of the phenomenon of oscillatory transformation of "robot-like" structures:** AFM scans of the imaging field shown in Figure 6 over a period of 14 hours. The time stamps are marked on each figure. The absolute value of time marked in minutes in each figure is the time that has elapsed since the previous image was acquired.

**Fig. S15.**

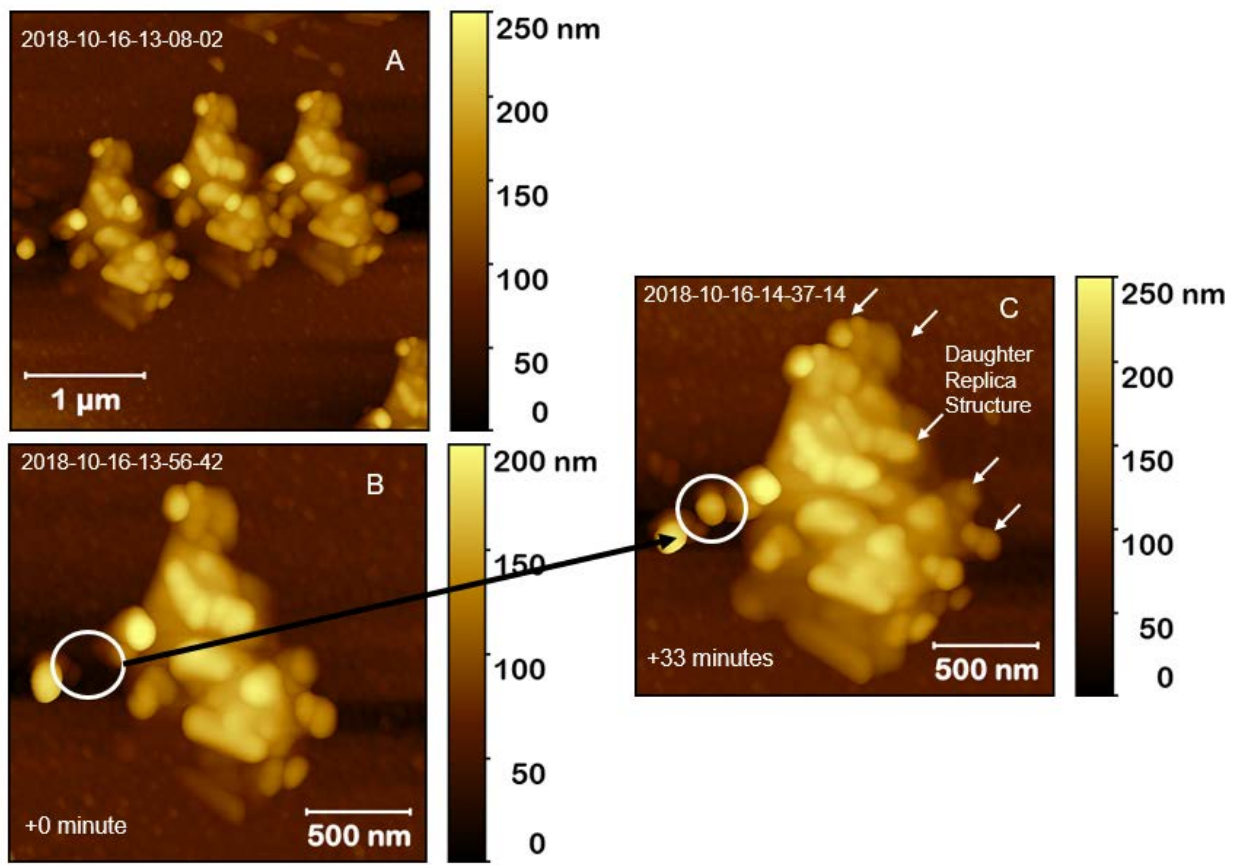

**Figure S15: Higher magnification AFM images of the complex 3-dimensional structures formed from the nano-particles:** In A we show three identical images pointing in a specific direction. In B, we show a higher magnification AFM image of an individual 3-dimensional structure and in C, we show replica copies of these structures. The white circle in B and C show an additional growth feature added in C.

**Fig. S16.**

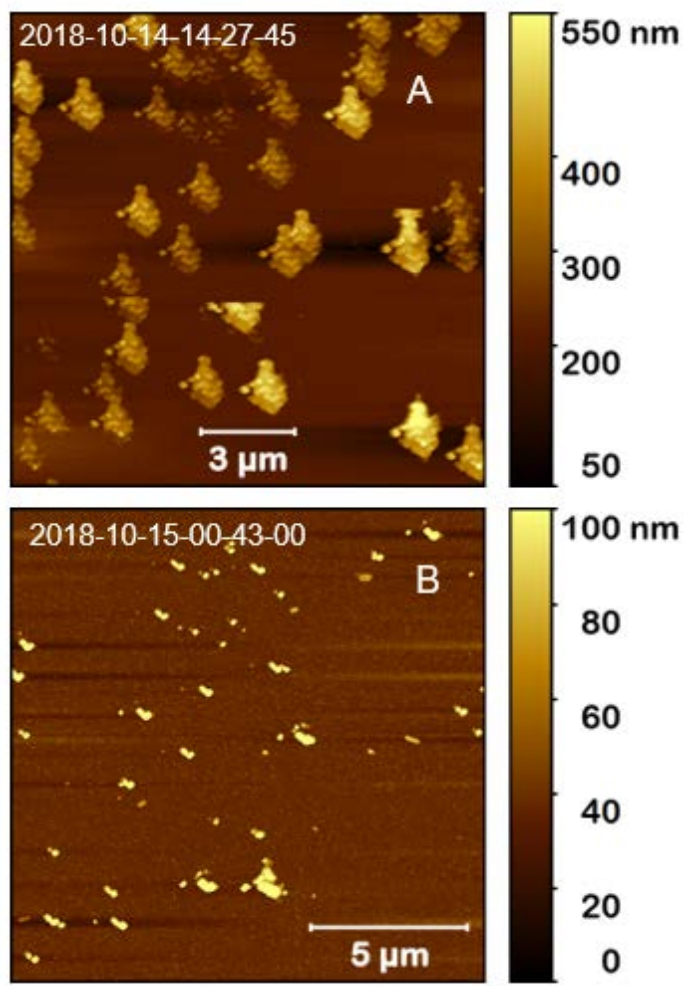

**Figure S16: The phenomenon of transformation between 3-dimensional structures and nano-particles**: The phenomenon was observed several hours earlier than that shown in Figure S14 and at a location that was 33.65 mm away: A is an AFM image of the 3-dimensional structures similar in form to those shown in Figure S14. B is the image of the same location after several hours showing the transformation back to nano-particles.

**Fig. S17:**

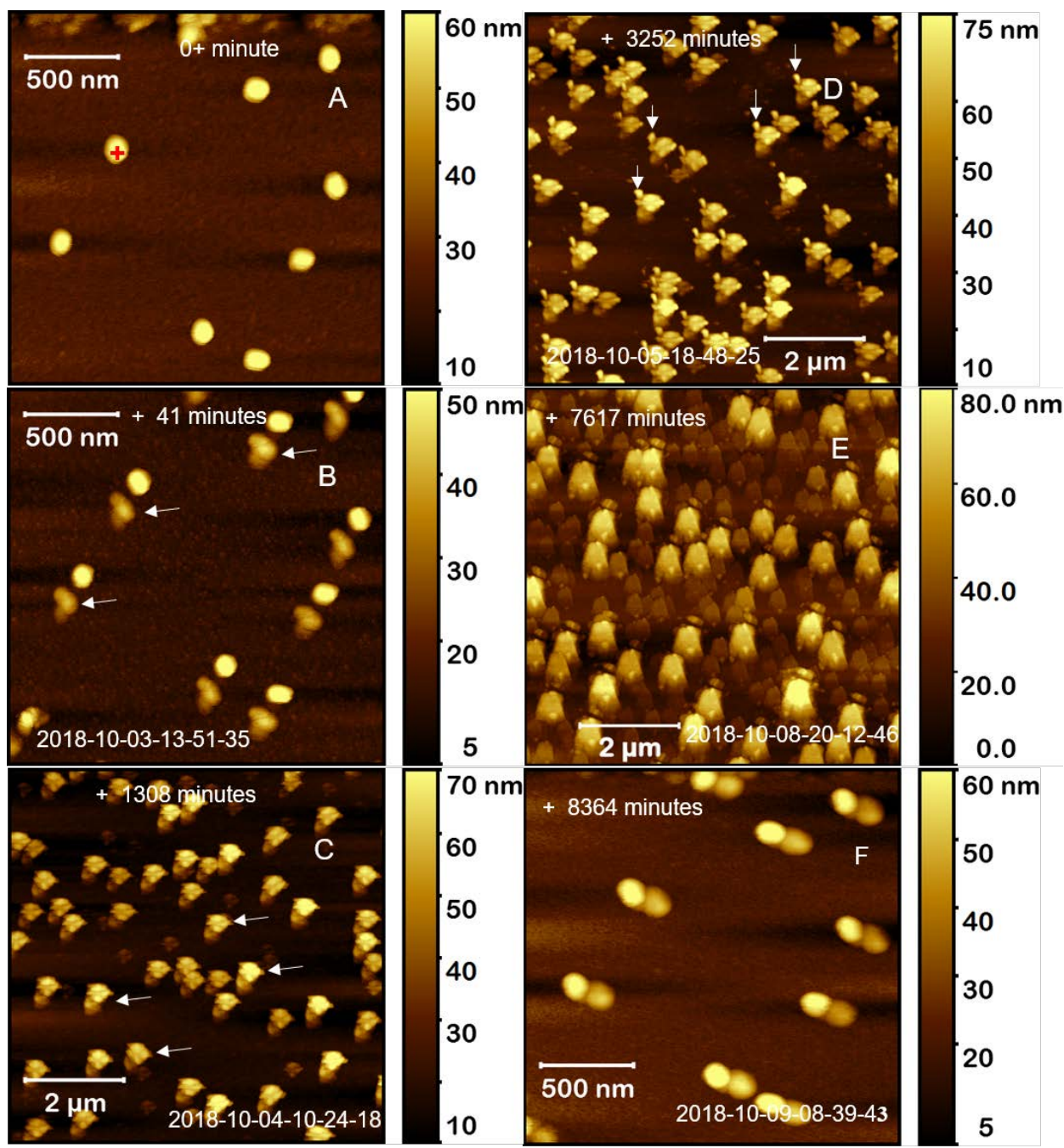

**Figure S17:** **Formation of novel 3-dimensional structures due to application of a 100nN force and the phenomenon of oscillatory transformation of the structure/forms:** In A we show the AFM image of the nano-particles prior to the application of a force. B is the image of the same location immediately after the application of the force, showing the creation of a 'daughter' particle by each 'mother' particle. All them are oriented in a specific direction. In C, we show the transformation of each nano-particle in B into an identical 3-dimensional structure with no particular orientation. In D, we show the transformation of each structure in C into a different well-defined identical 3-dimensional structure. All of them are oriented in a specific direction that is a mirror of that in B. In E, we show the transformation of each structure in D into another well-defined 3-dimensional structure with no particular orientation. In F we show transformation of the structures in E back to the original nano-particle structures. All the images have been date stamped.

**Fig. S18.**

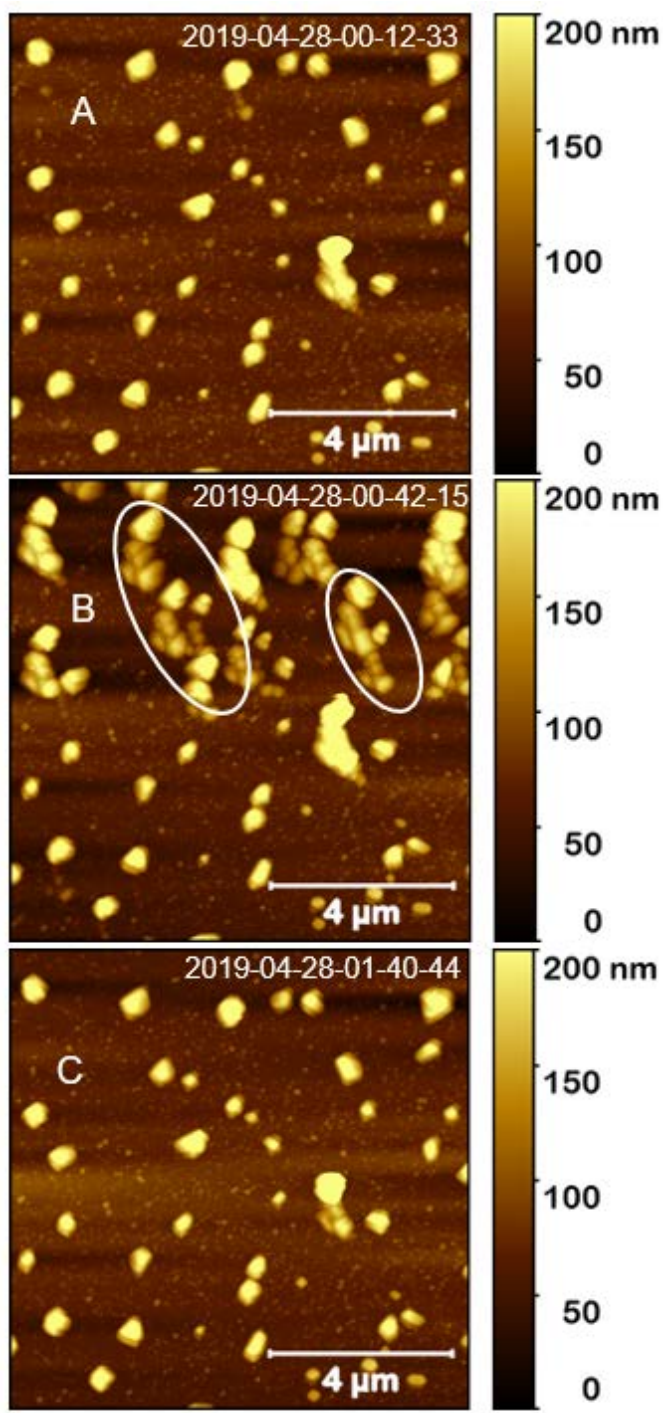

**Figure S18: Demonstration of the phenomenon of formation of complex 3-dimensional structures from nano-particles when subjected to a force and the oscillatory transformation between the two structures:** A is an AFM image prior to application of the force. B is the image of the same location immediately after the application of the force. C is the image of the same location after a few hours.

**Fig. S19:**

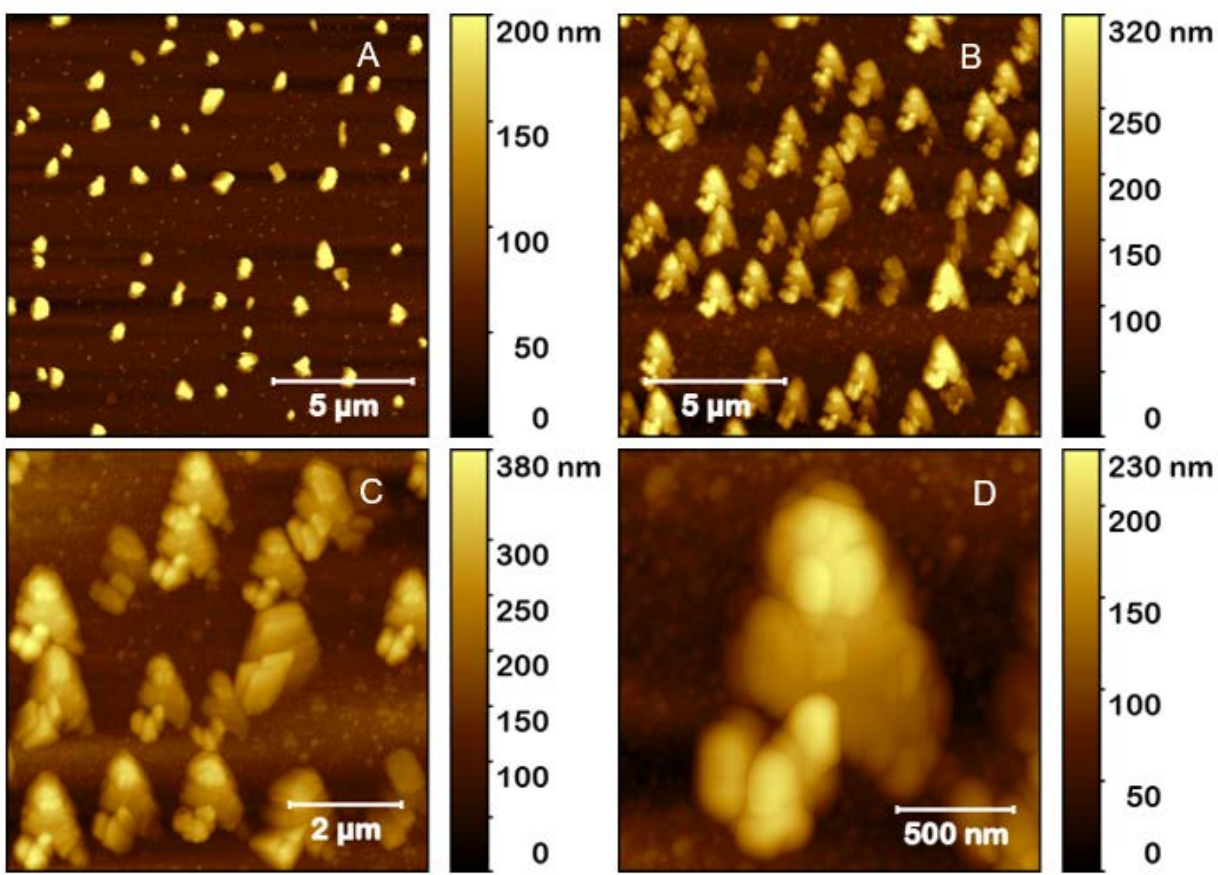

**Figure S19: Demonstration of the phenomenon of formation of complex 3-dimensional structures from nano-particles when subjected to a force:** A is an AFM image of the nano-particles prior to application of the force. B is the image of the same location immediately after the application of the force. C and D are magnified images of the 3-dimensional structures.

**Fig.S20**.

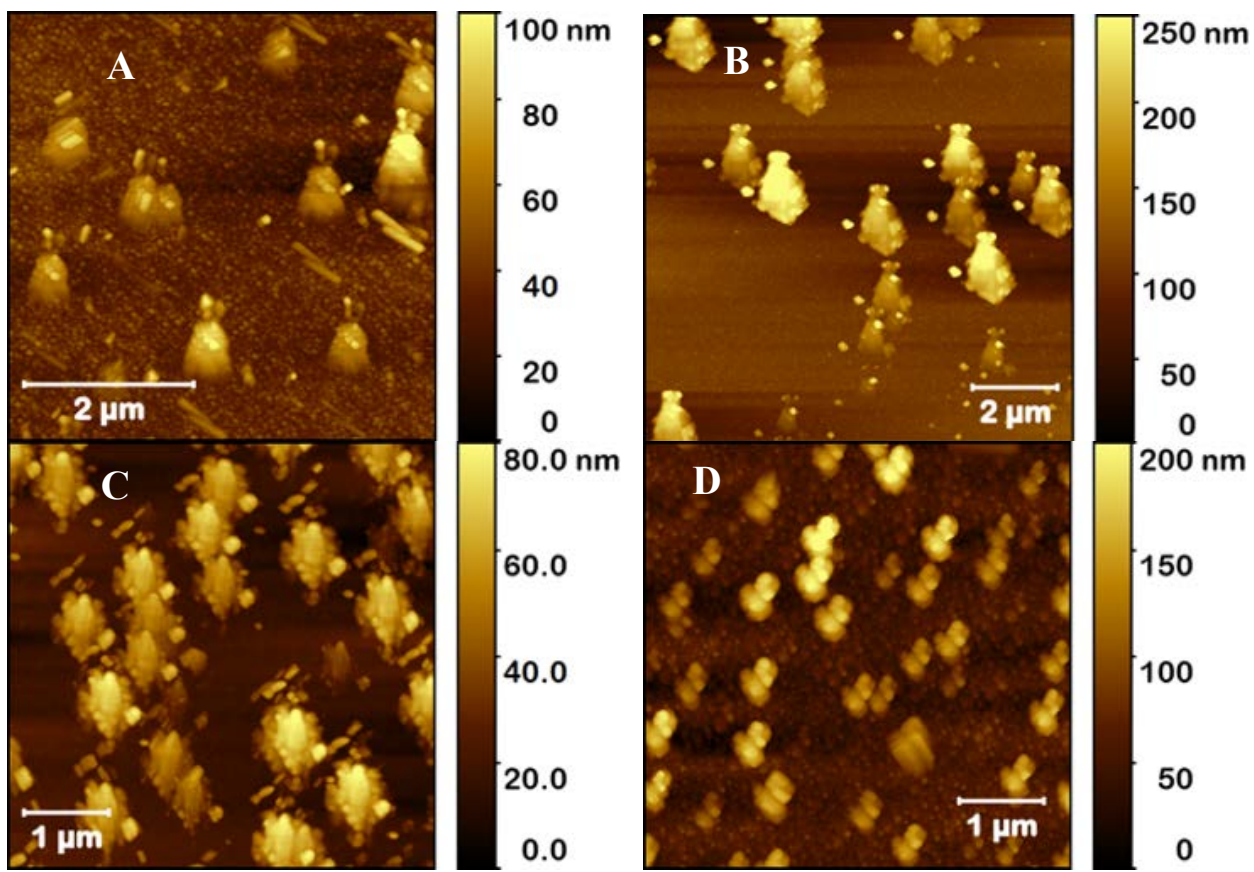

**Figure S20: Collection of different 3-dimensional structures created by nano-particles as a force response:** In A and B we show AFM images of 3-dimensional structures during a transient stage of their formation. C and D show two more examples of complex 3-dimensional structures created by nano-particles.